\documentclass[%
 reprint,
 showkeys,
 amsmath,amssymb,
 aps,
]{revtex4-2}
\usepackage[english]{babel}

\usepackage{graphicx}
\usepackage{dcolumn}
\usepackage{bm}
\usepackage{hyperref}
\usepackage{tcolorbox}
\usepackage{booktabs}
\usepackage{bm}

\DeclareUnicodeCharacter{2009}{\,}

\begin{document}

\preprint{APS/123-QED}

\title{Cell Competition Driven by Secreted Ligands:\\ Modeling Liver Metastasis of Colorectal Cancer}

\author{Hossein Nemati}
\affiliation{%
Institute for Theoretical Physics, Utrecht University, Princetonplein 5, 3584 CC Utrecht, The Netherlands
}%
\email{h.nemati@uu.nl}
\author{Saskia Jacoba Elisabeth Suijkerbuijk}
\affiliation{%
Division of Developmental Biology, Institute of Biodynamics and Biocomplexity, Department of Biology, Faculty of Science, Utrecht University, Padualaan 8, 3584 CH Utrecht, the Netherlands
}%
\author{Joost de Graaf}%
\affiliation{%
Institute for Theoretical Physics, Utrecht University, Princetonplein 5, 3584 CC Utrecht, The Netherlands~ 
}%





\begin{abstract}
Cell competition in multicellular organisms has been shown to play a critical role during the development of organisms, cancer progression, and in the establishment and maintenance of tissue homeostasis. Various mechanisms of cell competition have been identified, including active elimination via mechanical forces or induced apoptosis, as well as competition for nutrients and other beneficial factors.
A recent experiment demonstrated hallmarks of cell competition, associated with cell cycle dynamics, between liver progenitor cells and colorectal cancer cells [Krotenberg Garcia~\textit{et al.}, iScience~\textbf{27}, 109718 (2024)].~\nocite{Ana2024} However, a mechanistic explanation for this form of competition remains lacking.
Here, we present a mean-field model of competition for signaling ligands, coupled with cell cycle dynamics, to provide such an understanding.
Our model captures the salient features of the experiment, including population dynamics and cell cycle variations.
We demonstrate that secretion of a beneficial factor by cells, coupled with the enhanced uptake efficiency of cancer cells, suffices to reproduce the experimental outcome.
Our model, reminiscent of competition for secreted growth factors, provides insight into the minimal level of complexity required to achieve the observed competitive outcome as well as its link to cell cycle dynamics. It can also serve as a general framework for studying biological populations with 
growth-stage-dependent
competition over consumer-produced products.
\end{abstract}
\maketitle


\section{\label{sec:intro}Introduction}

Cell competition is an evolutionary process whereby cells with high survival and proliferative capacities, reflecting higher \textit{fitness}, are preferentially selected over less-fit cells~\cite{van_neerven_cell_2023, van_luyk_cell_2024}.
This process can occur through direct cell-cell interactions, secretion of biochemical factors, or mechanical forces~\cite{baker_emerging_2020, bowling_cell_2019, parker_cell_2021, van_luyk_cell_2024}, which in some cases can even result in the elimination of neighboring cells~\cite{krotenberg_garcia_active_2021, van_neerven_apc-mutant_2021, yum_tracing_2021, flanagan_notum_2021, parker_cell_2021}.
Cells may also compete for limited shared resources in an ecological manner~\cite{kedia_mehta_competition_2019, freischel_frequenc_dependent_2021, moreno_cells_2002, vincent_steep_2011, paluskievicz_t_2019, chang_metabolic_2015}, although this is not traditionally categorized in cell competition in the literature of cell biology.
The concept of cell competition was first introduced in 1974 by Morata and Ripoll, who observed disappearance of mutant cells when surrounded by wild types in the \textit{Drosophila} wing disc~\cite{morata_minutes_1975}. 
Since that time, the mechanisms of competition have been clarified through a series of studies on interactions between different cell types in \textit{Drosophila}~\cite{moreno_cells_2002,rodrigues_activated_2012,simpson_parameters_1979, simpson_differential_1981, johnston_drosophila_1999, baker_mechanisms_2017, soares_autocrine_2024} and more recently in mice~\cite{oliver_ribosomal_2004, oertel_cell_2006, claveria_myc-driven_2013, sancho_competitive_2013, ellis_distinct_2019, krotenberg_garcia_active_2021, yum_tracing_2021, Ana2024}. A historical review of cell competition studies was recently performed by Morata, himself~\cite{morata_cell_2021}.
%


Cell competition processes are generally classified into two categories: active and passive competition. In active competition, one cell type interferes with the behavior of another through mechanisms like induction of apoptosis~\cite{moreno_cells_2002, rodrigues_activated_2012, krotenberg_garcia_active_2021, ziosi_dmyc_2010, karim_ectopic_1998, neto-silva_evidence_2010, tyler_genes_2007}, differentiation~\cite{van_neerven_apc-mutant_2021, ellis_distinct_2019}, or cell cycle modifications~\cite{Ana2024, krotenberg_garcia_active_2021}. In contrast, passive cell competition is characterized by the absence of interference between cell types, although a selective advantage may still arise for one type over the other~\cite{van_luyk_cell_2024, van_neerven_cell_2023}, through for example, specific mutations~\cite{amoyel_neutral_2014, vermeulen_defining_2013, snippert_biased_2014} or position-dependent advantages~\cite{scheele_identity_2017}.

Cell competition is a double-edged sword. On the one hand, competition can serve as a quality control mechanism for the integrity of tissues in development or in homeostasis~\cite{kim_picking_2020, van_neerven_cell_2023}. On the other hand, certain types of cancer abuse competition to increase their proliferation capacity~\cite{van_luyk_cell_2024, bowling_cell_2019}. For a deeper biological insight into cell competition mechanisms, the reader is referred to recent comprehensive reviews in Refs.~\cite{van_neerven_cell_2023, van_luyk_cell_2024, baker_emerging_2020, bowling_cell_2019, parker_cell_2021}.

Numerous scientific studies have explored the modeling of cell competition. While a comprehensive review of all modeling approaches is still lacking, partial overviews can be found in Refs.~\cite{csikasz_nagy_cooperation_2013, cumming_toward_2024}.
Here, we briefly summarize the three major lines of modeling, in order to contextualize our work. The first category includes models based on ordinary differential equations (ODEs), which describe the dynamics of different populations and their essential (shared) resources using differential equations. 
The populations can be interpreted as cell types, which interact indirectly via resource competition, or directly, as is the case for predator-prey dynamics. Prominent examples of such models include \textit{MacArthur's model}~\cite{MacArthur1970, Chesson1990}; \textit{Tilman's model}~\cite{Tilman1980}; and the \textit{Lotka--Volterra model}, as well as its generalizations~\cite{wangersky_lotka-volterra_1978, cherniha_construction_2022, malcai_theoretical_2002}. For instance, the Lotka-Volterra model has been used to model competition among breast-cancer cell lines~\cite{freischel_frequenc_dependent_2021}. These models can also connect the cell cycle and competition for shared resources~\cite{chignola_ab_2007, de_la_cruz_coarse-graining_2017, cruz_stochastic_2016, ponce_bobadilla_age_2019}.

The second category of quantitative models originates from the fields of evolutionary dynamics and population genetics. These models are typically based on stochastic processes that govern population dynamics through discrete time steps~\cite{ewens_mathematical_2004, nowak_evolutionary_2006}. Prominent examples include the Moran~\cite{moran_random_1958} and Wright--Fisher~\cite{wright_evolution_1931, fisher_genetical_1930} models.
Buder and colleagues have used a simple Moran model to study tumor progression for a range of tissues~\cite{buder_patterns_2019}. 
These frameworks can also be extended to cover \textit{structured populations} using graphs, which opens the field \textit{evolutionary graph theory}~\cite{lieberman_evolutionary_2005}. Game theory is also used in these models to describe different types of competitive interactions between individuals~\cite{smith_evolutionary_1986, szabo_evolutionary_2007}. 
However, a limitation in this category is that many detailed biological properties are encoded in aggregated quantities such as fitness and game payoff, that do not distinguish between individuals of the same type.

The third category of quantitative models are cell-based models, also known as agent-based models.  These models consider cells as physical entities and describe their interactions using equations of motion based on potentials or forces.
Examples of these models are particle-based methods~\cite{akiyama_mathematical_2017, kirchner_vicsek-type_2024, camley_collective_2016, smeets_emergent_2016, szabo_phase_2006, li_competition_2022},  the vertex model~\cite{staple_mechanics_2010, alt_vertex_2017, lange_vertex_2025, bi_density-independent_2015, barton_active_2017}, Voronoi models~\cite{bi_motility_driven_2016, sussman_anomalous_2018, giavazzi_flocking_2018},  cellular Potts~\cite{glazier_simulation_1993, hogeweg_evolving_2000, boas_cellular_2018, szabo_cellular_2013, nemati_cellular_2024, braat_shape_2025}, and multi-phase-field models~\cite{nonomura_study_2012, moure_phase_field_2021, monfared_multi_phase_field_2025}. For the interested reader, these modeling approaches are comprehensively reviewed in Ref.~\citenum{alert_physical_2020}. These frameworks have been applied to study, for example, competition for resources~\cite{jiang_multiscale_2005}, and competitive interactions arising from the mechanical properties of cells~\cite{carpenter_physical_2024, brezin_mechanically_driven_2024, schoenit_force_2025}, cell motility~\cite{alsubaie_modelling_2024}, cell density effects~\cite{gradeci_cell_scale_2021}, and cell cycle dynamics~\cite{li_competition_2022, pak_mathematical_2024}.

In a recent experimental study by Krotenberg Garcia and colleagues~\cite{Ana2024}, cell competition was observed between liver progenitor cells and colorectal cancer cells, a finding relevant given that liver is the most common site of colorectal cancer metastasis~\cite{tang_nomogram_2021, engstrand_colorectal_2018, tsilimigras_liver_2021}. The observed competition manifested as asymmetric variations in proliferation rates of the two types compared to their pure conditions, as well as alteration in the distribution of liver cells across cell cycle phases~\cite{Ana2024}. Despite the experimental evidence of competition, the specific type of signaling or cell-cell interactions responsible for these observations remains unknown and is still under investigation.
In this work, inspired by the experimental observations~\cite{Ana2024} and motivated by the pursuit of a \textit{minimally complex} model, we propose a mean-field framework that captures the experimental results by considering competition for secreted signaling ligands. Modeling studies have explored cell competition and its interplay with the cell cycle in terms of active competition~\cite{pak_mathematical_2024, gradeci_cell_scale_2021}, 
mechanical competition~\cite{li_competition_2022, carpenter_physical_2024}, and 
competition for externally provided resources~\cite{cruz_stochastic_2016, de_la_cruz_coarse-graining_2017, ponce_bobadilla_age_2019}. However, scenarios involving competition for secreted ligands and their impact on cell cycle dynamics remain poorly characterized.
Our model provides a plausible mechanism linking this form of competition to cell cycle regulation within a minimal, unified framework. 
To the best of our knowledge, this is the first study to incorporate growth-stage-dependent competitive behavior, and consumer-produced resources dynamics in the literature of consumer-resource models. Our work can also be extended to ecological scenarios where similar competitive behavior is relevant.

\subsection*{\label{sec:exp_data_analysis}The experimental observations}

Krotenberg Garcia and colleagues~\cite{Ana2024} have recently studied cell competition between liver cells (either progenitor cells or hepatocytes) and colorectal cancer cells,
 in the context of colorectal cancer metastasis to the liver, 
using mixed murine organoids and microtissues. 
Here, we briefly review the results most relevant to our study, that is competition between liver progenitor cells and colorectal cancer cells. We refer the reader to Refs.~\citenum{Ana2024},~\citenum{krotenberg_garcia_generation_2021} and~\citenum{lamprou_protocol_2025} for details of the experiment.

\begin{figure}[h]
\centering
\includegraphics[width=0.49\textwidth]{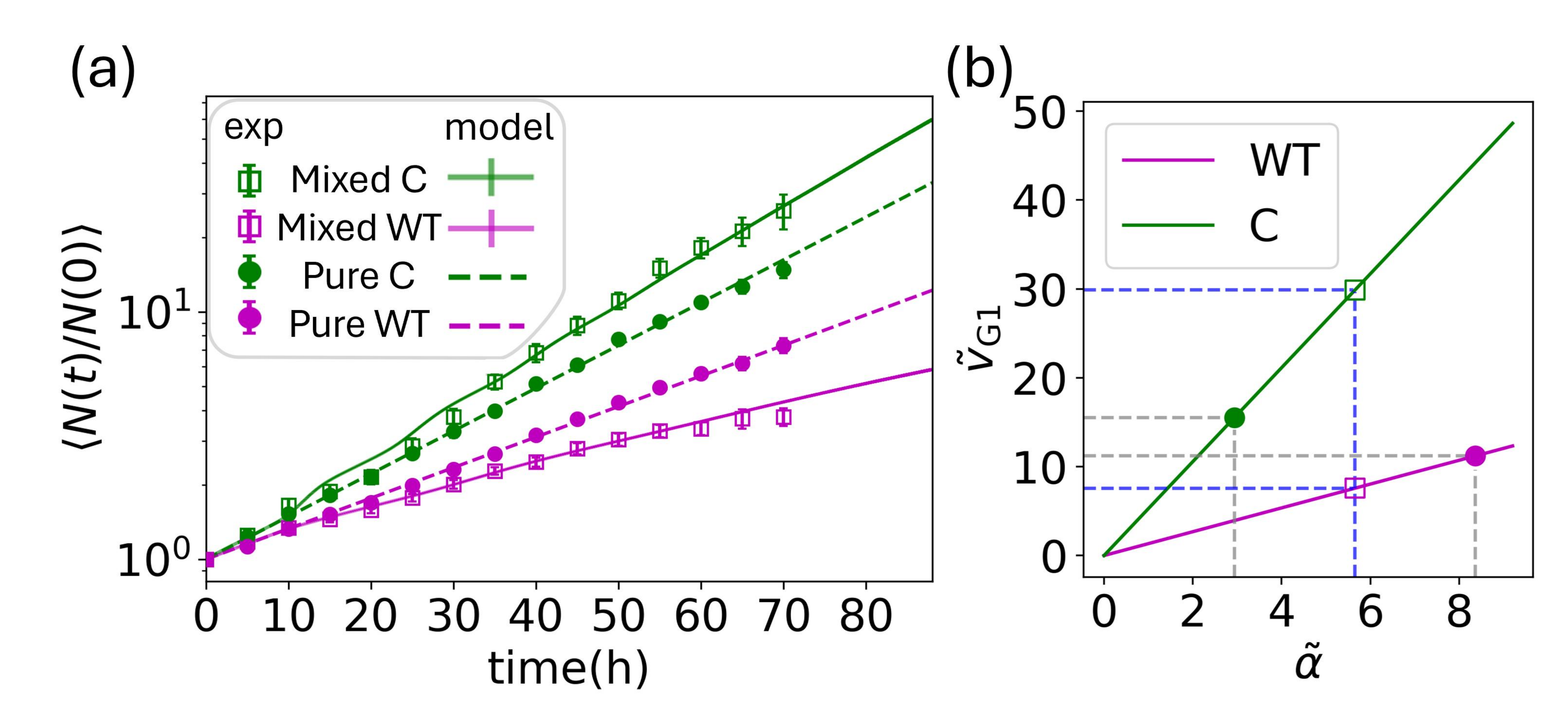}
\caption{Competition in terms of population growth. (a) Normalized population of wild-type and cancer cells are plotted. The curves are from the model, while the data points are the experimental data. (b) Progression speed in G1 phase, \(\tilde{v}_{\mathrm{G1}}\), as a function of growth factor concentration, \(\tilde{\alpha}\), for wild-type and cancer cells. The solid circular markers show the equilibrium points in the pure conditions. The open square markers show the initial values of \(\tilde{v}_{\mathrm{G1}}\) for wild-type and cancer cells in an example mixed organoid with a 50\,:\,50 composition. The gray and blue dashed lines serve to guide the eye. }
\label{fig:four_pop_vG1_prop}
\end{figure}

The most important result reported by Krotenberg Garcia~\textit{et al.}~\cite{Ana2024} is the population growth of the two cell types. The populations of wild-type and cancer cells in both \textit{pure} and \textit{mixed} organoids, were studied. In pure conditions, the populations of both cell types grew exponentially, with a higher rate for cancer cells. 
By performing a weighted least-square fitting on the experimental data to curves \(N(t) = N(0) \; e^{\beta t}\), we found $\beta_{\mathrm{W}} = 0.0284 \pm 0.002\; \text{h}^{-1}$ and $\beta_{\mathrm{C}} = 0.0398 \pm 0.002\; \text{h}^{-1}$ for pure wild-type and cancer populations, respectively.
Here, $\beta$ represents the (exponential) growth rate, i.e., fractional population increase per unit time. It is related to the population doubling time, $\tau_d$, through $\beta = \ln2/\tau_d$.
However, in mixed organoids, the growth of both cell types deviates from these exponential curves. Specifically, cancer cells in mixed organoids proliferate faster than in pure conditions, while the proliferation rate of wild-type cells decreases, as shown in Fig.~\ref{fig:four_pop_vG1_prop}(a). This clearly indicates a form of cell competition between wild-type and cancer cells in mixed organoids, mediated by interactions between the two cell types. However, as indicated previously, the precise mechanism underlying these interactions remains unknown.

The mixed organoids had a range of starting ratios, which revealed that the deviation from pure conditions in one type was exacerbated by a larger proportion of the other type. The experimental trend is reproduced in Fig.~\ref{fig:compos_model}, which will be discussed later. 
In addition to population growth, competition was shown to influence the dynamics of the cell cycle. For wild-type cells, competition altered the distribution of cells across different phases of the cell cycle. In the experiments~\cite{Ana2024}, using the FUCCI2 reporter~\cite{abe_visualization_2013}, the fraction of wild-type cells in G1 and those entering differentiation (G0 phase) increased in mixed organoids. To quantify the proportion of cells in the S/G2/M phases, EdU and pH3 labeling were also employed, and both methods consistently showed a reduction in the fraction of wild-type cells in these phases in mixed organoids compared with pure organoids. These trends are reproduced in Fig.~\ref{fig:phase_percentages}, and will be discussed further in the context of our modeling.

\section{\label{sec:model}A mean-field model for the cell competition}

We focus on identifying the simplest competition scenario that can account for all the above experimental observations while relying on biologically reasonable assumptions. We propose that cells compete for \textit{one} common environmental resource. However, this resource cannot be nutrients alone.
If cancer cells were simply \textit{superconsumers} of nutrients responsible for depletion in mixed organoids, the depletion would occur in pure cancer population as well, resulting in a detectable fall in the slope of their growth, which is not observed in the experiment~\cite{Ana2024}.
Instead, both pure populations exhibit sustained exponential growth throughout the experimental time window. This leads us to conclude that the pure populations grow in a self-sufficient manner. Specifically, if a resource is needed for population growth, it must be produced by the cells themselves to maintain the exponential growth. This points toward competition for autocrine/paracrine signaling ligands. In this mode of signaling, cells secrete molecules such as growth factors, which diffuse through the surrounding environment and act on neighboring cells as well as the secreting cell itself. Secreted growth factors, in turn, promote cell proliferation. Liver progenitor cells~\cite{kong_growth_2023, yang_sonic_2008, santoni_rugiu_progenitor_2005, salomon_transforming_1990, michalopoulos_hepatostat_2017, araacute_growth_2010}
and colorectal cancer cells~\cite{yarden_untangling_2001, sizeland_anti_sense_1992, lamonerie_igf_2_1995} are known to secrete growth factors as autocrine or paracrine signaling to support their own proliferation.

\subsection*{Model Formulation}
Our modeling approach is as follows. For each cell, we represent the position in the cell cycle by a quantity called 
the `cell phase' as a function of time, which we denote by $\phi(t)$. The value of $\phi$ ranges from $0$ 
(for a newborn cell) to $2\pi$ (for a cell about to divide). Without loss of generality, we may assign the interval $[0,\pi]$ to the G1 phase and $[\pi,2\pi]$ to the remaining phases 
of the cell cycle, comprised of phases S, G2, and M phases, which we denote by subscript `S' for simplicity. After a cell divides and passes $\phi=2\pi$, its daughter cells 
restart cycling with $\phi = 0$. We denote the progression speeds in G1 and in the rest of the cycle by $v_{\text{G1}}$ and $v_{\text{S}}$, respectively.

Instead of tracking single cells in the phase domain, we look at the $\phi$-space density of cells $\rho^{(\text{W,C})}(\phi, t)$, where W and  C denote wild-type and cancer cells. 
The total populations of wild-type and cancer cells are indicated using $W(t)$ and $C(t)$, respectively, which are found by 
\(W(t) = W_\text{D}(t) +\int_{0}^{2\pi}\rho^\mathrm{W}(\phi,t)d\phi\)
 and 
\(C(t)=\int_{0}^{2\pi}\rho^\mathrm{C}(\phi,t)d\phi\).
We denote the number of differentiated wild-type cells using $W_\text{D}(t)$. The differentiation process for liver progenitor cells in culture, takes several days~\cite{Ana2024}. Here, `differentiated' refers to cells that have already entered the G0 phase and have initiated the differentiation process. In the experiments~\cite{Ana2024}, transition into the G0 phase is unidirectional and necessarily leads to differentiation. Therefore, the terms `G0 state' and `differentiated state' are used interchangeably.

It is accepted that cycling cells undergo regulatory commitment to either exit the cell cycle by entering the G0 phase or continue progressing, prior to reaching a specific point in mid-to-late G1 phase~\cite{Blagosklonny2002, PLANASSILVA1997, Pardee1974, johnson_start_2013, pardee_animal_1978, ezhevsky1997, zetterberg1995}. This point is known as the `restriction point' (R-point). Beyond the R-point, cells are committed to proceed through the subsequent stages of the cell cycle. In this work, we assume the R-point is at the end of the G1 phase. We introduce a transition rate, \(r_\mathrm{D}\), representing the rate at which cells exit G1 and enter the G0 state for differentiation. This rate governs the differentiation of cells prior to reaching the R-point. Naturally, once cells have passed the R-point, no such differentiation occurs.

We make several reducing assumptions to make the model tractable. In summary, these are as follows: 
(i) Each cell secretes a growth factor with a constant rate, $S$. This can be dependent on the cell type ($S_\mathrm{W}$ and $S_\mathrm{C}$), but here we assume them to be equal.
(ii) Cells consume the growth factor at a rate proportional to its concentration in the environment. However, the proportionality factor for the cancer cells is higher than that for the wild-type cells. This means that, at the same concentration of growth factor, cancer cells consume it at a higher rate than wild-type cells. This assumption is supported by the fact that colorectal cancer cells~\cite{spano_epidermal_2005}, as other cancer cells of epithelial origin~\cite{yarden_untangling_2001} are known to overexpress growth factor receptors on their membranes. Thus, if we denote the growth factor concentration with [GF], the consumption rates for the wild-type and cancer cells are \(\mu_{\text{W}}\text{[GF]}\) and \(\mu_{\text{C}}\text{[GF]}\), respectively, where \(\mu_{\text{C}} > \mu_{\text{W}}\) are proportionality factors.
(iii) To avoid geometrical complications, we approximate the concentration of the growth factor by its total amount per capita, \(\alpha(t) = M(t)/N(t)\), where \(M(t)\) and \(N(t)\) are the instantaneous values for the total mass of the growth factor and the total population, respectively.

Growth factors play a significant role in the regulation of the G1 phase~\cite{wang_regulation_2021, foster_regulation_2010, hulleman_regulation_2001,lukaszewicz_contrasting_2002}, which includes modification of G1 length~\cite{lukaszewicz_contrasting_2002, molinie_cortical_2019}. Therefore, as the next assumption, (iv) the speed of cell cycle progression in the G1 phase, $v_{\text{G1}}$, is considered sensitive to the growth factor consumption rate.  Specifically, \(v_{\text{G1}}^{\text{W}}\) and \(v_{\text{G1}}^{\text{C}}\) are assumed to be proportional to the growth factor consumption rates:
\begin{equation}
v_{\text{G1}}^{(\text{W}, \text{C})}(t)
\;=\;F_0\,\mu_{(\text{W}, \text{C})}\,\alpha(t).
\label{eq:v_G1_W_C}
\end{equation}
The factor \(F_0\) is a constant that converts the units of consumption rate (\(\mathrm{kg/h}\)) to those of progression speed (\(\mathrm{h}^{-1}\)). 
(v) Unlike G1, the progression speed, $v_{\text{S}}$, is constant in the phases S/G2/M and independent of growth factor concentration. This is supported by the presence of an R-point in the cell cycle, and by the observation that phases other than G1 show less variability in length~\cite{icard_interconnection_2019, dong_cyclin_2018}.

A key assumption (vi) underlying the use of a mean-field model, is that the secreted substance rapidly becomes a homogeneously distributed `common good' on time scales much shorter than those of cell division. 
It is easy to justify this assumption by estimating the time scale of diffusion of secreted factors. The diffusion coefficient of growth factors can be estimated using the Stokes–Einstein relation,    $D = k_{\mathrm{B}} T / (6 \pi \eta r)$. Growth factors are typically proteins with molecular sizes ($r$) on the order of a few nanometers. Assuming a temperature of \( T = 310\;\text{K} \) and the viscosity of water (\(\eta \approx0.001\;\text{Pa.s}\)), the diffusion coefficient for growth factors falls in the range of approximately \( 1 \times 10^{-11} \) to \( 1 \times 10^{-10}\;\text{m}^2/\text{s} \). Considering the length scale of organoids to be \( L \approx 200\;\mu\text{m} \), this leads to a diffusion time scale $\tau \sim {L^2}/(6D)$ in the range of \( 100 \) to \( 1000\;\text{s} \).
This provides an estimate of the diffusion time, which may be longer in the presence of macromolecules, yet is still expected to remain much shorter than the typical cell-division time ($\sim$20 hours).

\begin{figure}[h]
\centering
\includegraphics[width=0.49\textwidth]{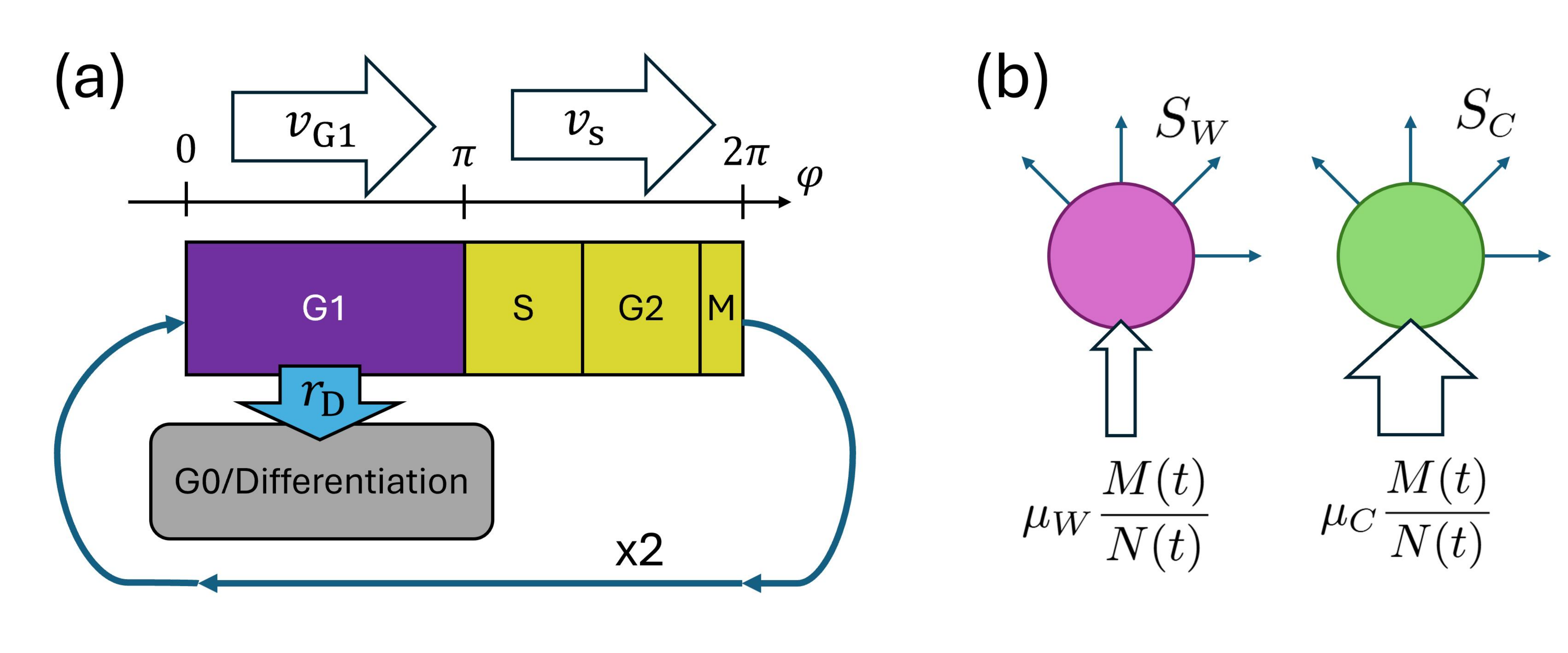}
\caption{Graphical summary of the modeling approach. (a) Schematic representation of the cell cycle used in our mean-field modeling. The G1 phase is shown in purple, and the rest of the cell cycle in yellow. The gray box indicates the wild-type cells that have exited their cycle to start the differentiation process. Differentiation can happen from any point in G1, as indicated using the blue arrow. The axis indicates how the cell phase $\phi$ changes as a cell progresses through G1 with speed $v_{\text{G1}}$; and through S, G2, and M phases with speed $v_{\text{S}}$. Each time a cell exits the M phase ($\phi = 2\pi$), it is doubled and both daughters reenter the G1 phase with $\phi = 0$. (b) Visualization of the assumptions underlying the secretion and the uptake rates in the mean-field model. The wild-type (magenta) and cancer cells (green) are shown as well as their secretion rates indicated by $S_\mathrm{W}$ and $S_\mathrm{C}$, respectively. The uptake rates are proportional to the total amount of growth factor per capita with proportionality factors \(\mu_\mathrm{W}\) and \(\mu_\mathrm{C}\) for wild-type and cancer cells where \(\mu_\mathrm{C}\) is assumed to be greater than \(\mu_\mathrm{W}\), as indicated using the width of the arrow.  }
\label{fig:assumptions_schematic}
\end{figure}

Lastly, in the experiments, no apoptosis was observed. 
Therefore, as the assumption (vii), we do not include any apoptosis or death rate term in the model. 
The model assumptions for the cell cycle and the secretion and consumption processes are graphically illustrated in Fig.~\ref{fig:assumptions_schematic}.

Under the above assumptions, the evolution equations for the wild-type and cancer cells read 
\begin{equation}
    \frac{\partial \rho^{(\text{W,C})}}{\partial t} = -\frac{\partial}{\partial \phi}\left(\rho^{(\text{W,C})} v^{(\text{W,C})}\right)- r^{(\text{W,C})} \rho^{(\text{W,C})},
    \label{eq:general_continuity}
\end{equation}
which represent an advection-reaction system. The differentiation rate $r$ is zero for the cancer cells for all $\phi$, while for the wild-type cells it is constant ($r_{\text{D}}$) for $\phi < \pi$ and zero otherwise. 
The progression speed, \(v^{(\text{W,C})}\), should be replaced by \(v^{(\text{W,C})}_\text{G1}\) for \(\phi<\pi\) and by \(v_\text{S}\) for \(\phi>\pi\). The boundary conditions for Eq.~\ref{eq:general_continuity}, are given by flux continuity at $\phi = \pi$, and doubling at $\phi = 2\pi$, where the division occurs:
\begin{equation}
    v_{\mathrm{G1}}^{(\mathrm{W}, \mathrm{C})}(t)\,\rho^{(\mathrm{W}, \mathrm{C})}(\pi^-,t)
    = v_{\mathrm{S}}\;\rho^{(\mathrm{W}, \mathrm{C})}(\pi^+,t),
    \label{eq:BC_cont}
\end{equation}
and
\begin{equation}
v_{\mathrm{G1}}^{(\mathrm{W}, \mathrm{C})}(t)\,\rho^{(\mathrm{W}, \mathrm{C})}(0,t)
= 2\,v_{\mathrm{S}}\,\rho^{(\mathrm{W}, \mathrm{C})}(2\pi,t).
\label{eq:BC_doubl}
\end{equation}
For the total amount of the growth factor, $M(t)$, which according to Eq.~\ref{eq:v_G1_W_C} determines $v_{\mathrm{G1}}^{(\mathrm{W}, \mathrm{C})}$, we obtain:
\begin{align}
\frac{d M}{dt}
&=\;\Bigl(S - \mu_{\text{W}} \,\tfrac{M}{N}\Bigr)\,W(t)
\;+\;\Bigl(S - \mu_{\text{C}} \,\tfrac{M}{N}\Bigr)\,C(t),
\label{eq:M}
\end{align}
where \(N(t) = W(t) +C(t)\), is the total population of the organoid.
Finally, for the population of wild-type cells going into differentiation, \(W_\text{D}(t)\), we have:
\begin{equation}
    \frac{d W_\text{D}}{dt}
=\;r_\text{D} \int_{0}^{\pi} \rho^{\text{W}}(\phi, t)\,d\phi.
\label{eq:W_D}
\end{equation}

In Eqs.~\ref{eq:v_G1_W_C}--\ref{eq:W_D},
the parameters \(r_{\mathrm{D}}, v_{\mathrm{S}}, F_0, \mu_{\mathrm{W}}, \mu_{\mathrm{C}}\), and \(S\) appear. To simplify the system and reduce the number of parameters, we nondimensionalized the equations. We used \(1/\beta_{\mathrm{W}}\) and \(1/F_{0}\) as the characteristic time and mass, respectively. 
Hence, using a tilde to denote dimensionless quantities, we define
\(\tilde{t}=\beta_{\mathrm{W}}t,~\tilde{M} = F_0 M,~\tilde{r}_{\mathrm{D}}={r_{\mathrm{D}}}/{\beta_{\mathrm{W}}},~\tilde{v}_\mathrm{S}={v_\mathrm{S}}/{\beta_{\mathrm{W}}},~\tilde{\mu}={\mu}/{\beta_{\mathrm{W}}},~\text{and}~\tilde{S}={S}F_0/{\beta_{\mathrm{W}}}\), 
where 
\(\mu\) 
carry a subscript either W or C, indicating  wild-type or cancer cells, respectively.
Therefore, we can express 
Eqs.~\ref{eq:v_G1_W_C}--\ref{eq:W_D}
in dimensionless form, for which we provide solutions.

\subsection*{Solution for pure conditions}
For pure conditions of both types, we assume that the distribution of cells in \(\phi\) is already in quasi-steady state. This means for both types, \(\partial(\rho(\phi,\tilde{t})/{N(\tilde{t})}) / \partial \tilde{t} = 0\), and for wild-type cells, the differentiated fraction, \(f_\text{D} = W_\text{D}/W\) is constant, i.e., \(\partial(W_\text{D}/W) / \partial \tilde{t}=0\). 
The solution for the pure wild-type population is:
\begin{equation}
\rho^{\mathrm{W}}(\phi,\tilde{t}) = \left\{
\begin{array}{ll}
  \rho^{\mathrm{W}}(0,0)\,\exp\!\left(-\frac{1 + \tilde{r}_{\mathrm{D}}}{\tilde{\mu}_{\mathrm{W}} \tilde{\alpha}_{\mathrm{W}}}\,\phi\right) \,e^{\tilde{t}}&\text{, } \phi \in(0,\pi), \\
  \rho^{\mathrm{W}}(\pi^+,0) \,\exp\!\left(-\frac{\phi - \pi}{\tilde{v}_{\mathrm{S}}}\right) \,e^{\tilde{t}}  & \text{, } \phi \in(\pi,2\pi);
\end{array}
\right.
\label{eq:rho_pure_W_sol}
\end{equation}
\begin{equation}
    W_\mathrm{D}(\tilde{t}) =  f_\mathrm{D}\;W(0)\;e^{\tilde{t}};
    \label{eq:W_D_pure_W_sol}
\end{equation}
\begin{equation}
    \tilde{M}(\tilde{t}) = 
    \tilde{\alpha}_\mathrm{W}
    \;W(0)\;e^{\tilde{t}},
    \label{eq:M_pure_W_sol}
\end{equation}
where $\tilde{\alpha}_\mathrm{W}=\tilde{S}/\left(1+\tilde{\mu}_\mathrm{W}\right)$ is the equilibrium concentration for pure wild-type population. 
However, the boundary conditions~\ref{eq:BC_cont} and~\ref{eq:BC_doubl}, together with conditions $\int_{0}^{2\pi}\rho^{\mathrm{W}}(\phi,0)d\phi = (1-f_\mathrm{D})W(0)$,  $dW_\mathrm{D}/d\tilde{t} = \tilde{r}_\mathrm{D}\int_{0}^{\pi}\rho^{\mathrm{W}}(\phi,\tilde{t})d\phi$, and the fact that in experiments $f_\mathrm{D}\approx0.2$, reduce the problem to two free parameters, $\tilde{S}$ and $\tilde{v}_\mathrm{S}$. The interdependencies among our equations allow us to obtain the other quantities, see the supplement.
\begin{equation}
    \tilde{r}_\mathrm{D} = f_\mathrm{D}/\left(2-f_\mathrm{D} - e^{\pi/\tilde{v}_\mathrm{S}}\right);
\end{equation}
\begin{equation}
    \tilde{\mu}_\mathrm{W} = 1/\left( \tilde{S} m/\left({1+\tilde{r}_\mathrm{D}}\right) -1\right);
\end{equation}
\begin{equation}
    \rho^{\mathrm{W}}(0,0) = 
    2m(1-f_\mathrm{D})W(0)/\left(1+\tilde{r}_\mathrm{D}\left(e^{\pi/\tilde{v}_\mathrm{S}}-1\right)\right);
\end{equation}
and
\begin{equation}
    \rho^{\mathrm{W}}(\pi^+,0) = \rho^{\mathrm{W}}(0,0)(\tilde{\mu}_\mathrm{W} \tilde{\alpha}_\mathrm{W}/\tilde{v}_\mathrm{S}) \,e^{-m\pi},
\end{equation}
where for simplicity, we introduce $m = ({1 + \tilde{r}_{\mathrm{D}}})/({\tilde{\mu}_{\mathrm{W}} \tilde{\alpha}_{\mathrm{W}}}) = \ln2/\pi-1/\tilde{v}_\mathrm{S}$. Thus, the solution for pure wild-type population is complete once \(\tilde{S}\) and \(\tilde{v}_\mathrm{S}\) are given.

For pure cancer cells, where differentiation is assumed absent, we have:
\begin{equation}
\rho^{\mathrm{C}}(\phi,\tilde{t}) = \left\{
\begin{array}{ll}
  \rho^{\mathrm{C}}(0,0)\,\exp\!\left(-\frac{\gamma}{\tilde{\mu}_{\mathrm{C}} \tilde{\alpha}_{\mathrm{C}}}\,\phi\right) \,e^{\gamma \tilde{t}}&\text{, } \phi \in(0,\pi); \\
  \rho^{\mathrm{C}}(\pi^+,0) \,\exp\!\left(- \frac{\gamma(\phi - \pi)}{\tilde{v}_{\mathrm{S}}} \right) \,e^{\gamma \tilde{t}}  & \text{, } \phi \in(\pi,2\pi);
\end{array}
\right.
\label{eq:rho_pure_C_sol}
\end{equation}
\begin{equation}
    \tilde{M}(\tilde{t}) = \tilde{\alpha}_\mathrm{C} \;C(0)\;e^{\gamma \tilde{t}},
    \label{eq:M_pure_C_sol}
\end{equation}
where $\gamma=\beta_\mathrm{C}/\beta_\mathrm{W}$, and $\tilde{\alpha}_\mathrm{C} = \tilde{S}/\left( \gamma+\tilde{\mu}_\mathrm{C}\right)$ is the equilibrium concentration for pure cancer cells.
Again, assuming \(\tilde{S}\) and \(\tilde{v}_\mathrm{S}\) are given, we use boundary conditions~\ref{eq:BC_cont} and~\ref{eq:BC_doubl} as well as \(\int_{0}^{2\pi}\rho^{\mathrm{C}}(\phi,0)d\phi = C(0)\), to obtain \(\tilde{\mu}_\mathrm{C}\), \(\rho^{\mathrm{C}}(0,0)\) and \(\rho^{\mathrm{C}}(\pi^+,0)\):
\begin{equation}
    \tilde{\mu}_\mathrm{C} =\gamma^2/ \left(\tilde{S}q-\gamma\right);
\end{equation}
\begin{equation}
    \rho^{\mathrm{C}}(0,0) =
    2 q C(0);
\end{equation}
and
\begin{equation}
    \rho^{\mathrm{C}}(\pi^+,0) =  C(0)\,\left({\gamma/{\tilde{v}_\mathrm{S}}}\right) \,e^{\gamma\pi/\tilde{v}_\mathrm{S}},
\end{equation}
where, for brevity, we define $q = \gamma / \left(\tilde{\mu}_\mathrm{C}\tilde{\alpha}_\mathrm{C}\right)  = \ln2/\pi-\gamma/\tilde{v}_\mathrm{S}$.
Since we are interested in the normalized
populations, we set $W(0)=1$ and $C(0)=1$ for pure conditions.
Populations calculated from densities above, are plotted in Fig.~\ref{fig:four_pop_vG1_prop}(a) as well as the experimental data. 
As is obvious from the equations, in the quasi-steady state, cells are exponentially distributed in \(\phi\), and the densities at any \(\phi\) increase exponentially with time, reproducing the same growth behavior as we see in the experiment. 
Constant values of growth factors per capita ($\tilde{\alpha}_\mathrm{W}$ and $\tilde{\alpha}_\mathrm{C}$) here, arise from the quasi--steady-state assumption.
Full derivations are presented in the SI.

\subsection*{Solution for mixed conditions}
The solutions we previously presented for pure organoids guarantee the exponential growth shown in Fig.~\ref{fig:four_pop_vG1_prop}(a) and ensure that the fraction of differentiating cells remains equal to \(f_\mathrm{D}\) in pure conditions, regardless of the values of \(\tilde{S}\) and \(\tilde{v}_\mathrm{S}\).
For the mixed organoids, we numerically solve Eq.~\ref{eq:general_continuity}  for both types, as well as Eqs.~\ref{eq:M} and~\ref{eq:W_D} using the boundary conditions provided in  Eqs.~\ref{eq:BC_cont} and~\ref{eq:BC_doubl}. The initial cell counts for mixed organoids are determined by drawing random samples from synthetic distributions fitted to experimental data. 
Specifically, we fitted exponential distributions in the form
\(f(n)=(1/\theta)\; e ^{-(n-n_0)/\theta}\)
to the initial cell numbers in the experiments, and sampled randomized values to serve as initial conditions for the model.
For wild-type and cancer cells, the distribution parameters turned out to be 
\(\theta_{\text{W}}=53.0,~n_0^{\text{W}}=14.0, ~,\theta_{\text{C}}=38.2~\text{and}~n_0^{\text{C}}=2.0\).
Full details of the distribution fitting process are provided in the SI. 
From these distributions, 500 pairs of random samples were drawn to serve as the initial counts of each cell type in mixed organoids.
Once the initial counts of wild-type and cancer cells are established, the initial concentration of growth factors is computed as a weighted average of the equilibrium concentrations, which is
\begin{equation}
\tilde{M}(0) = N(0)\;\tilde{\alpha}_{\text{avg}} =  {W(0)\;\tilde{\alpha}_\mathrm{W}+C(0)\;\tilde{\alpha}_\mathrm{C}}.
\label{eq:M_init_cond}
\end{equation}
Thus far, $\tilde{S}$ and $\tilde{v}_\mathrm{S}$ are assumed given.
To fully solve the problem, these parameters must therefore be identified. Their values are obtained by minimizing a cost function that penalizes the discrepancy between the model predictions and the experimental data. 
The cost function comprises three terms which account for the populations in mixed organoids, the statistics of wild-type cells, and the effects of initial percentage of cells.
It is expressed as: 
\begin{equation}
    L(\tilde{S}, \tilde{v}_\mathrm{S}) = 
    \omega_\text{p}
    L_{\text{pop}} + 
    \omega_\text{s}
    L_{\text{stat}} + 
    \omega_\text{i}
    L_{\text{init}}.
    \label{eq:cost_function}
\end{equation}
The terms \(L_{\text{pop}}\), \(L_{\text{stat}}\), and \(L_{\text{init}}\)  represent the errors for population dynamics, cell statistics in different phases, and the effect of initial composition, respectively. They are the weighted squared errors between the experimental data and the predictions of the model.
The factors \(\omega_\text{p}\), \(\omega_\text{s}\), and \(\omega_\text{i}\) are used to weight different terms and, for simplicity, are all considered \(1/3\).
Full details on calculation of each term are presented in the SI. 

\section{\label{sec:results}Results}
In this section, we first present the model predictions obtained with the best‑fit parameter set and compare them with the experimental observations. The independent parameters were found to be \(\tilde{S}=19.6\pm0.2\) and \(\tilde{v}_{\mathrm{S}}=10.8\pm0.1\).
The biophysical meanings of these numerical values follow from their definitions in the Model Formulation section: $\tilde{S}=S/S_0$ represents the ratio of the secretion rate to a reference mass rate, $S_0=\beta_\mathrm{W}/F_0$. Here, $\beta_\mathrm{W}$ denotes a rate ($\beta_\mathrm{W}=\ln2/\tau^\mathrm{W}_d$), while $1/F_0$ defines growth factor consumption per unit phase progression, as given by Eq.~\ref{eq:v_G1_W_C}.
If we denote the total growth factor consumption during the G1 phase by $\Delta M_\mathrm{G1}$, we obtain $S_0=(\ln2/\pi)\Delta M_\mathrm{G1}/\tau_{d}^\mathrm{W}$.
In addition, $\tilde{v}_\mathrm{S}$ directly yields $v_\mathrm{S}=\beta_\mathrm{W}\tilde{v}_\mathrm{S}$, which in turn gives the duration of the S/G2/M phases as $T_\mathrm{S} = \pi/v_\mathrm{S} =10.1\;\mathrm{h}$.
Further interpretation will be provided in the Discussion section.
From these, the remaining parameters followed \(\tilde{\mu}_{\mathrm{W}}=1.34\pm 0.05\), 
\(\tilde{\mu}_{\mathrm{C}}=5.24\pm 0.4\), 
\(\tilde{\alpha}_{\mathrm{W}}=8.37\pm 0.3\), 
\(\tilde{\alpha}_{\mathrm{C}}=2.94\pm 0.2\), 
\(\tilde{r}_{\mathrm{D}}=0.43\pm 0.005\).
A comparison between $\tilde{\mu}_\mathrm{W}$
and $\tilde{\mu}_\mathrm{C}$
indicates that the growth factor consumption capacity of cancer cells is approximately four times greater than that of wild-type cells.
Using the obtained numerical values, the densities $\rho^{(\mathrm{W,C})}(\phi, t)$ and the total populations $W(t)$ and $C(t)$ were found. The resulting  populations overlay the experimental data from Ref.~\cite{Ana2024} for both pure and mixed organoids, as shown in Fig.~\ref{fig:four_pop_vG1_prop}(a). In pure conditions, the growth curves are exponential, as expected from the analytical solutions. 
In the mixed conditions, the growth rate of wild-type cells decreases, whereas that of cancer cells increases. 

The variations in the populations shown in Fig.~\ref{fig:four_pop_vG1_prop}(a), can be explained by panel~(b) of the same figure. In this panel, we have plotted the dimensionless G1-progression speed, $\tilde{v}_{\mathrm{G1}}$, as a function of dimensionless growth factor per capita, $\tilde{\alpha}$, for wild-type and cancer cells,  as described in the original form, by Eq.~\ref{eq:v_G1_W_C}. The curve for cancer cells has a higher slope due to their higher uptake capacity. 
This allows cancer cells to have a higher G1 progression speed than wild-type cells, not only at any equal concentration of growth factor $\alpha$, but also at lower values  (down to $(\mu_\text{W}/\mu_\text{C})\alpha$). It can be understood by comparing the equilibrium concentrations of pure organoids, shown by solid circular markers in Fig.~\ref{fig:four_pop_vG1_prop}(b). 
In mixed organoids, the growth factor concentration lies between the two extremes observed in pure wild-type and pure cancer populations. An example value is shown by open square markers in Fig.~\ref{fig:four_pop_vG1_prop}(b). Notably, the different  slopes for  wild-type and cancer cells
in Fig.~\ref{fig:four_pop_vG1_prop} (b) 
lead to an asymmetry in how 
$\tilde{v}_{\mathrm{G1}}$ varies for the two types. That is, mixing exposes wild-type cells to a lower concentration than in their pure conditions, resulting in a reduced $\tilde{v}_{\mathrm{G1}}$. In contrast, cancer cells are exposed to a higher concentration than in pure conditions, leading to an increase in $\tilde{v}_{\mathrm{G1}}$. 
This characteristic consequently leads to the asymmetric variations in populations, which is shown in Fig.~\ref{fig:four_pop_vG1_prop}(a). There is, however, a slight mismatch between the model and the experiment for cancer cells in mixed organoids in the interval \(10\sim30 \; \textrm{h}\). 
This is because, for a number of the mixed organoids, the population of cancer cells was observed to decrease in this interval.

\begin{figure}[]
\centering
\includegraphics[width=0.49\textwidth]{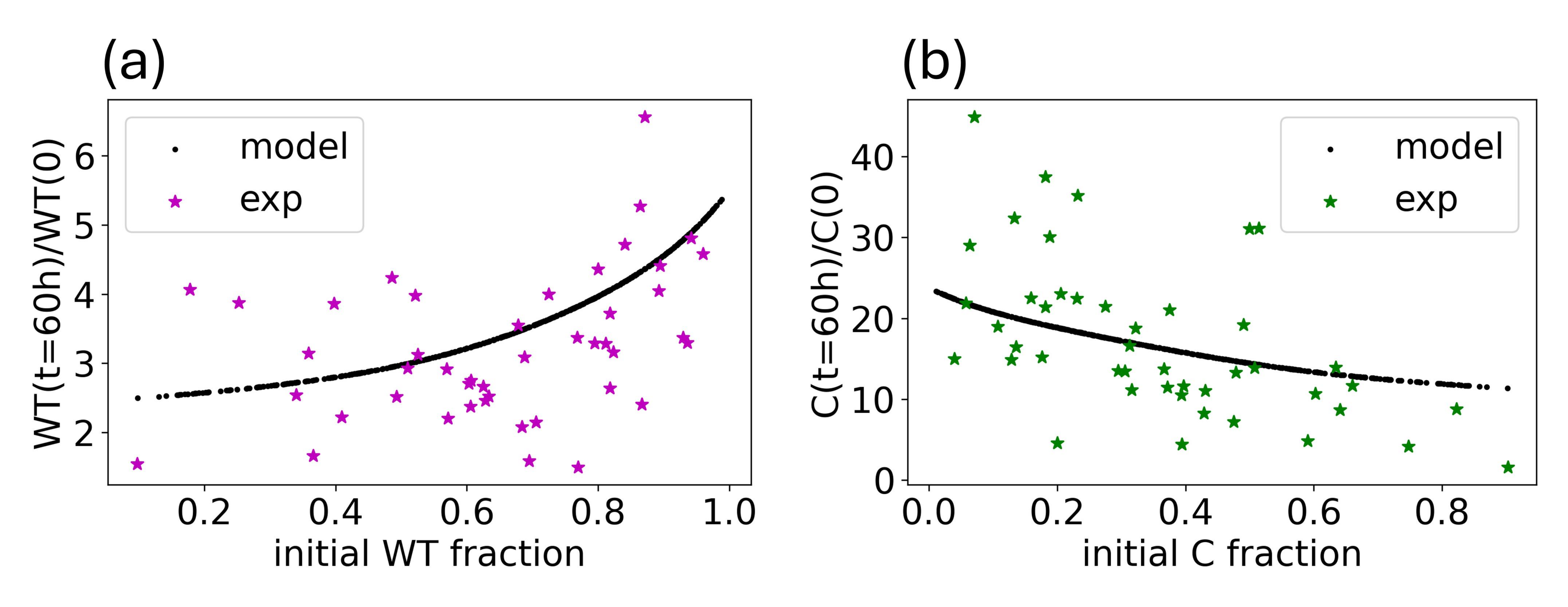}
\caption{Normalized population growth of each cell type at $t=60$ h vs. their initial fraction in the mixture. The data are shown for (a) wild-type cells and (b) cancer cells.  The stars show the data from the experiment, while the dots from the model. The number of organoids for the experimental data is \(n=45\) and for the model, \(n=500\).  }
\label{fig:compos_model}
\end{figure}

Next, we considered the impact of the initial composition of the organoid.
Following the experimental workflow, we plotted the normalized population size of each cell type at \(t = 60\,\mathrm{h}\) versus its initial fraction in the mixed organoids. The resulting data of the model, together with the experimental data, are shown in Fig.~\ref{fig:compos_model}.
We capture the same trend of the experiments: the growth of wild-type cells increases with their starting proportion, whereas the growth of cancer cells decreases as their own starting proportion rises. However, the match between the trend in model and experiment is better for wild-type cells than for cancer cells.  

\begin{figure}[h]
\centering
\includegraphics[width=0.49\textwidth]{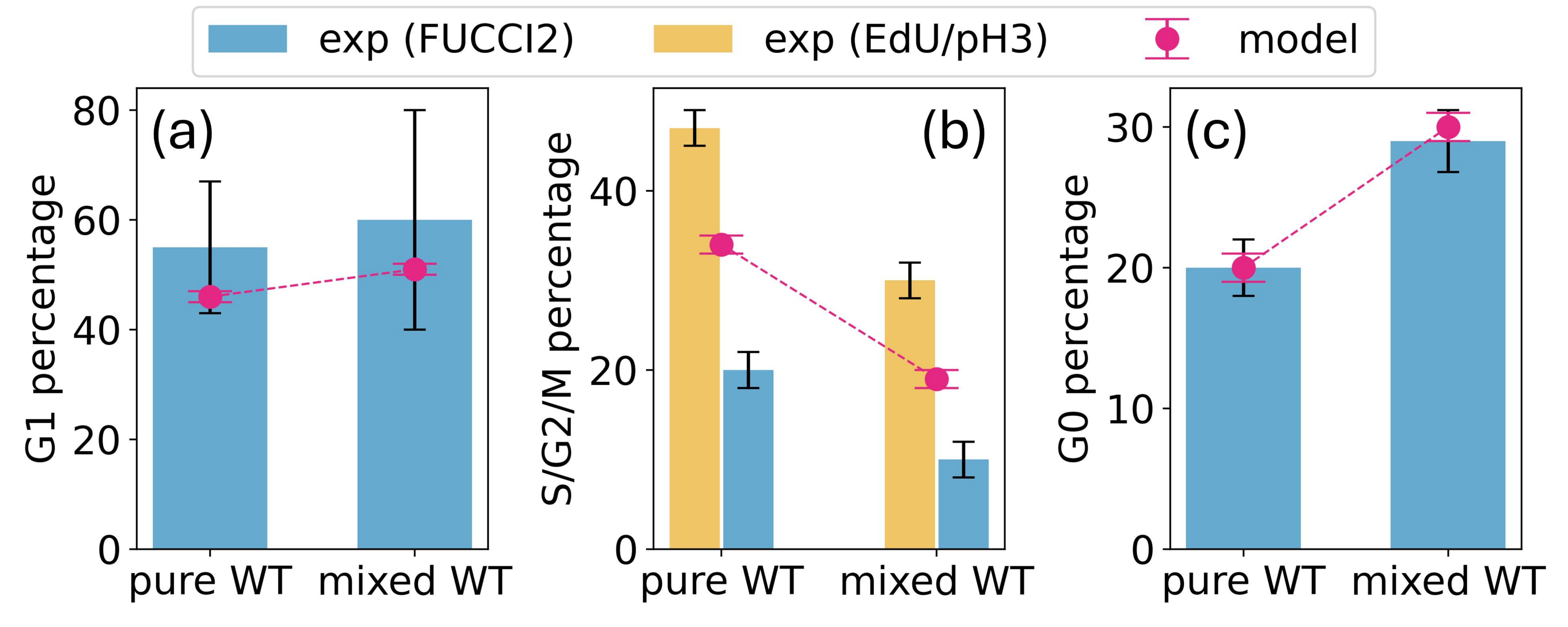}
\caption{Comparison of experimental data and model predictions for the proportions of wild-type cells in (a) G1 phase, (b) S/G2/M phases, and (c) G0 phase of the cell cycle. The data are provided for pure and mixed organoids at \(t=60\;\mathrm{h}\). The experimental data are taken from Ref.~\citenum{Ana2024}. 
For the cells in S/G2/M phases, the two different experimental measurement methods are shown.
Uncertainties for experimental data indicate SEM. For the model, a basic 1\% uncertainty due to rounding was considered. The statistical SEM was smaller. The dashed lines connect the data of the model to guide the eye.}
\label{fig:phase_percentages}
\end{figure}

Competition alters the distribution of wild-type cells across cell cycle phases, as observed experimentally. 
We have made a comparison of the percentages of cells in different phases between model and experiment. Figure~\ref{fig:phase_percentages} presents the percentages of wild-type cells in each phase of the cell cycle obtained from the experiment and the model at \(t = 60\;\mathrm{h}\), the only time point for which experimental data are available.  
The model reproduces the proportions G0 and G1 within the corresponding experimental uncertainties.
For the S/G2/M phases, two complementary assays were employed, as described in the section on experimental observations: FUCCI2 reporter and combined EdU + pH3 protocol.
Averaging these two measurements yields a balanced reference value which was used as the target value in the cost minimization.  This is what the model reproduces, as shown in the middle panel of Fig.~\ref{fig:phase_percentages}. 
Most importantly, the mean-field formulation captures every experimentally observed shift in phase distribution that occurs when wild-type cells reside in mixed organoids compared to the pure conditions. As a complementary analysis, we studied the model’s predictions for the temporal evolution 
of phase-specific cell fractions, which is discussed in SI.

\begin{figure}[h]
\centering
\includegraphics[width=0.49\textwidth]{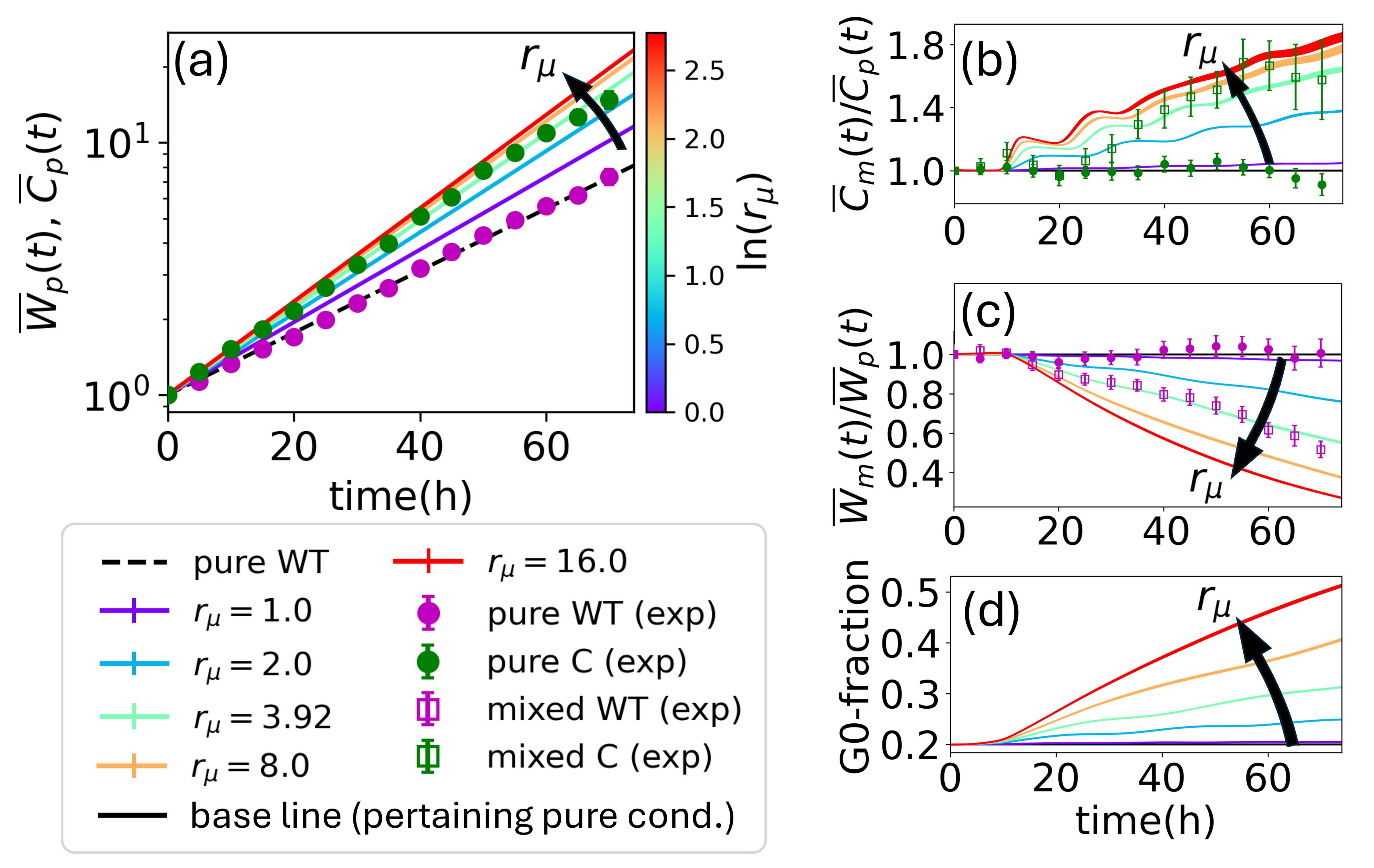}
\caption{Influence of uptake coefficient ratio on competition. 
(a) Normalized populations under pure conditions for both cell types at different values of $r_\mu = \tilde{\mu}_\mathrm{C} / \tilde{\mu}_\mathrm{W}$. 
(b, c) Ratios of normalized populations in mixed and pure conditions for (b) cancer and (c) wild-type cells as a function of time. 
(d) Fraction of differentiated wild-type cells (G0 phase) as a function of time for different values of $r_\mu$.  Colors in all panels correspond to the color bar in panel~a.
The arrows show the direction along which $r_\mu$ increases. 
The legend applies to all panels.}
\label{fig:mu_variations}
\end{figure}

To evaluate the robustness of the model, we studied the sensitivity of its predictions to the parameters \(\tilde{S}\) and \(\tilde{v}_\mathrm{S}\).
This was done by solving the problem for a range of values of the two parameters, which covered almost \(\pm 5\%\) of the fitted values.
We first ensured that the captured critical values for \(\tilde{S}\) and \(\tilde{v}_\mathrm{S}\) define a global minimum in the landscape of cost function, by analyzing the Hessian matrix.
Next, we found the directions defining the highest and the lowest curvature of the cost function using eigenvectors of the Hessian matrix, which were
\(\tilde{S} -\tilde{S}^{\text{fit}} = 0.32(\tilde{v}_\text{S}-\tilde{v}^{\text{fit}}_\text{S}) \) and 
\(\tilde{S} -\tilde{S}^{\text{fit}} = -3.1(\tilde{v}_\text{S}-\tilde{v}^{\text{fit}}_\text{S}) \), 
respectively.
This shows that the direction for the least sensitivity aligns closer to the direction of \(\tilde{S}\), while sensitivity is higher in the direction of \(\tilde{v}_\mathrm{S}\). 
The reason for this anisotropy is that the variation in \(\tilde{v}_\mathrm{S}\) controls the fitting process in two ways. First, it determines the fractions of cells in S/G2/M, which in turn contributes to \(L_\text{stat}\) in the cost function. Furthermore, the fractions of division time between the S/G2/M phases and the G1 phase are controlled by \(\tilde{v}_\mathrm{S}\). Since the G1 phase is the window for competition, \(\tilde{v}_\mathrm{S}\) also indirectly controls population growth, contributing to \(L_\text{pop}\). In contrast, the model shows less dependence on the secretion term, \(\tilde{S}\). This means that the model predictions would remain qualitatively unchanged with slightly different secretion rates and even with unequal secretion rates for wild-type and cancer cells. 
See the Sensitivity analysis section in the Supplementary Information for a detailed discussion.
The difference between the uptake coefficient factors $\tilde{\mu}_\mathrm{W}$ and $\tilde{\mu}_\mathrm{C}$ 
is a key factor resulting in the asymmetric variation of growth curves shown in Fig.~\ref{fig:four_pop_vG1_prop}(a). The parameter fitting yields a ratio
\(r_{\mu} = {\tilde{\mu}_\mathrm{C}}/{\tilde{\mu}_\mathrm{W}} = 3.92\) for the coefficients. 
To investigate how this ratio influences the competition, we varied $r_\mu$ by adjusting $\tilde{\mu}_\mathrm{C}$ while keeping all other parameters, including $\tilde{\mu}_\mathrm{W}$, fixed. The ratio $r_\mu$ is swept from $1$ to $16$, which encompasses the fitted value of $3.92$, and the resulting model behavior is shown in Fig.~\ref{fig:mu_variations}. 
Firstly, the normalized populations of wild-type and cancer cells under pure conditions are shown in Fig.~\ref{fig:mu_variations}(a). Since $\tilde{\mu}_\mathrm{W}$ is kept constant, the pure wild-type population remains unchanged, whereas the pure cancer population increases with \(r_\mu\). Even at \(r_\mu=1\), a difference exists between cancer and wild-type cells: a fraction of wild-type cells continually initiate differentiation and cease proliferating, while cancer cells are assumed not to differentiate at all. 
In the next step, we examined the \textit{level of competition} by considering two quantities. The first is the ratio of the normalized populations in mixed and pure conditions for each type, which indicates how strongly population growth deviates from the pure conditions as a result of competitive interactions. The second is the differentiated fraction of wild-type cells, representing the proportion of wild-type cells that have exited the cell cycle.
In panels~(b) and~(c) of Fig.~\ref{fig:mu_variations}, we plot the ratios $\overline{W}_\mathrm{m}(t) / \overline{W}_\mathrm{p}(t)$ and $\overline{C}_\mathrm{m}(t) / \overline{C}_\mathrm{p}(t)$, where $\overline{W}$ and $\overline{C}$ denote the normalized populations of wild-type and cancer cells, and the subscripts \emph{m} and \emph{p} refer to mixed and pure conditions, respectively. As shown in Figs.~\ref{fig:mu_variations}(b) and~(c), this ratio increases for cancer cells and decreases for wild-type cells as $r_\mu$ increases. Finally, Fig.~\ref{fig:mu_variations}(d) shows the fraction of differentiated wild-type cells  in mixed conditions over time for different values of $r_\mu$. The results indicate that higher ratios of uptake coefficients accelerate cell cycle exit in wild-type cells. Clearly, competitive interactions intensify as the ratio of uptake coefficients between the two cell types increases.

\section{\label{sec:discussion}Discussion}

The physical meaning of \(\tilde{v}_{\text{S}}\) is directly related to cell division time which can be determined under pure conditions:
\begin{equation}
    T_\text{d}^\text{(W,C)} = \pi{\beta_{\text{W}}}^{-1}\left(1/{\tilde{v}_{\text{G1}}^{\text{(W,C)}}} + 1/{\tilde{v}_{\text{S}}}\right),
\end{equation}
where \({\tilde{v}_{\text{G1}}^{\text{(W,C)}}}\) is evaluated at the quasi--steady-state concentrations for each type, that is, \({\tilde{\mu}_{\text{(W,C)}}}{\tilde{\alpha}_{\text{(W,C)}}}\). 
We find $T_\text{d,pure}$ to be  $20.1\;\text{h}$ for wild-type cells and $17.4\;\text{h}$ for cancer cells. 
In mixed conditions, 
as is shown in Fig.~\ref{fig:four_pop_vG1_prop}(b), we always have \(\tilde{\alpha}_\mathrm{C}<\tilde{\alpha}<\tilde{\alpha}_\mathrm{W}\). This consequently defines an upper limit for \(T_\text{d}^\text{W}\), and a lower limit for 
\(T_\text{d}^\text{C}\), which are:
\begin{equation}
    T_\text{d,max}^\text{W} =\pi{\beta_{\text{W}}}^{-1}\left( \left(\tilde{\mu}_{\text{W}}\tilde{\alpha}_{\text{C}}\right)^{-1} + \tilde{v}_{\text{S}}^{-1}\right) = 38.3\;\text{h};
\end{equation}
and 
\begin{equation}
    T_\text{d,min}^\text{C} = \pi\beta_{\text{W}}^{-1}\left(\left(\tilde{\mu}_{\text{C}}\tilde{\alpha}_{\text{W}}\right)^{-1} + \tilde{v}_{\text{S}}^{-1}\right) = 12.8\;\text{h}.
\end{equation}
These two, respectively, describe situations where a wild-type cell or a cancer cell resides in an \textit{infinite} background of the other type. 
The calculated extreme values of the division time, together with those obtained under pure conditions, define the intervals within which the division times under mixed conditions lie. For wild-type cells, the division time in mixed conditions is always longer than that in pure conditions, whereas for cancer cells it is always shorter. In other words, we always have $T_\text{d,pure}^\text{W}<T_\text{d,mixed}^\text{W}<T_\text{d,max}^\text{W}$ and 
$T_\text{d,min}^\text{C}<T_\text{d,mixed}^\text{C}<T_\text{d,pure}^\text{C}$.

The duration of G1 phase is shown to be a determinant in cell cycle timing~\cite{dong_cyclin_2018} and correlates with cell fate~\cite{lukaszewicz_contrasting_2002}. Hence, it is informative to 
determine the fraction of cell division time dedicated to the G1 phase, which in general is:
\begin{equation}
    g = {T_{\text{G1}}}/{T_{\text{d}}} = \left(\pi/\tilde{v}_\text{G1}\right)/\left(\pi/\tilde{v}_\text{G1} + \pi/\tilde{v}_\text{S}\right),
\end{equation}
where \(\tilde{v}_\text{G1}\) for pure wild-type and cancer cells is \(\tilde{\mu}_\text{W} \tilde{\alpha}_\text{W}\) and \(\tilde{\mu}_\text{C} \tilde{\alpha}_\text{C}\), respectively.
We obtain $g_\text{W,pure} = 0.49$ and $g_\text{C,pure} = 0.41$ for wild-type and cancer  cells, respectively. Given that G1 phase duration ratio is highly variable, spanning from  $0.1\sim0.15$ in pluripotent stem cells~\cite{boward_concise_2016} to more than \(0.6\) in different human cells~\cite{blank_scaling_2018}, the values we obtain here appear to be reasonable. Similarly to the division time, we can establish an upper bound for $g$ in wild-type cells, and a lower bound in cancer cells. These correspond to $T_\text{d,max}^\text{W}$ and $T_\text{d,min}^\text{C}$, yielding  $g^\mathrm{W}_\text{max} =0.73$ and $g^\mathrm{C}_\text{min} =0.20$, respectively. 
In the limiting case of long-term exposure, the average division time of wild-type cells approaches 
$T_\text{d,max}^\text{W}$, while that of cancer cells converges to their intrinsic value, 
$T_\text{d,pure}^\text{C}$. This occurs because, under these conditions, the cancer population 
dominates and $\alpha$ tends toward $\alpha_\mathrm{C}$. 
In addition, the elongation of G1 phase in wild-type cells leads to an increase in differentiation rate.

Autocrine secretion rate of growth factors in epithelial cells, can be variable depending on the type of growth factors. For example, it can range from the order \(\sim 10^3\) molecules/h for TGF-\(\alpha\) to almost \(10^4 \sim 10^5\) molecules/h for EGF, the uptake rate being of the same order of magnitude~\cite{dewitt_quantitative_2001, joslin_egf_receptor_mediated_2007}. In our model, the uptake-to-secretion ratios for wild-type and cancer cells in their pure conditions are 
$\tilde{\mu}_\text{W}\tilde{\alpha}_\text{W}/\tilde{S} \approx 0.57$ and 
$\tilde{\mu}_\text{C}\tilde{\alpha}_\text{C}/\tilde{S} \approx 0.78$, respectively, 
with the higher value in cancer cells reflecting their enhanced proliferative capacity.
In addition, our sensitivity analysis showed that the model’s behavior is robust to variations in the secretion rate $\tilde{S}$.

The asymmetric variations in growth curves of the two types are the most insightful feature of the observed competition, which emerged from biased uptake capacities in the model. Our modeling choice of this is based on traits previously reported in the literature, including autocrine/paracrine signaling in liver cells~\cite{kong_growth_2023, yang_sonic_2008, santoni_rugiu_progenitor_2005, salomon_transforming_1990, michalopoulos_hepatostat_2017, araacute_growth_2010} and cancer cells~\cite{yarden_untangling_2001, sizeland_anti_sense_1992, lamonerie_igf_2_1995}, overexpression of growth factor receptors in cancer cells~\cite{spano_epidermal_2005, yarden_untangling_2001}, and the regulatory role of growth factors in the G1 phase of the cell cycle~\cite{wang_regulation_2021, foster_regulation_2010, hulleman_regulation_2001,lukaszewicz_contrasting_2002,lukaszewicz_contrasting_2002, molinie_cortical_2019}. The unperturbed exponential growth of pure organoids, as well as the initial composition dependence that is reproduced as natural outcomes of our choices, suggests that our model offers a plausible explanation of the experimentally observed competition.

The G1 phase in cell cycle is the stage where cells integrate division-commitment signals, and its duration is regulated by growth factor concentration~\cite{wang_regulation_2021, dalton_linking_2015, lukaszewicz_contrasting_2002, molinie_cortical_2019}.
Motivated by these findings, we assumed that the progression speed of cells in G1, \(v_{\mathrm{G1}}\), is—in the simplest case—linearly dependent on their growth factor uptake rates, which in turn implies proportionality to the growth factor concentration. Using this assumption, we reproduced the variations in cell phase distributions observed in wild-type cells. To keep the model minimal, we assumed the simplest dependence of progression speed on cell-cycle phase: a linear dependence on growth factor uptake in G1 (Eq.~\ref{eq:v_G1_W_C}), and a constant speed in S/G2/M phases ($v_\mathrm{S}$). Yet, this was sufficient to capture variations in both growth curves and cell-phase distributions. 
We also considered a step function-like averaging for the initial condition of \(\tilde{M}\), based on Eq.~\ref{eq:M_init_cond}. 
This leads to transient oscillations in the phase fractions in time, which are discussed in the SI.
Using the methods employed by Krotenberg Garcia and colleagues~\cite{Ana2024}, it was not possible to track these phase fractions in real time, as the measurements were terminal and thus precluded further observation of the same samples. It would be valuable to apply alternative, non-terminal approaches to evaluate these fractions at multiple time points, enabling a more robust assessment of the model's predictions.

Our model incorporates two key features: (i) biased competition for autocrine/paracrine signaling ligands and (ii) G1-phase sensitivity to those signals. In the broader consumer–resource framework, these translate to (i) competition on consumer‐produced resources and (ii) growth-phase-dependent competitive behavior.
Although each feature has been examined separately---consumer-produced resources as a model of agriculture~\cite{picot_apparent_2019} or as metabolic byproducts~\cite{cui_diverse_2021,cherniha_construction_2022}, and cell-phase-specific competition~\cite{pak_mathematical_2024, cruz_stochastic_2016, carpenter_physical_2024, de_la_cruz_coarse-graining_2017}, to the best of our knowledge, the first study to integrate both features in a single model. Our approach is readily applicable to any competitive community where consumers \textit{produce and consume} resources, and where their growth response depends on their age or proliferative status. 
Traditionally, in modeling competition using resource-consumer models, the process of reproduction is considered as a black box which appears as a time-derivative of population in the equations, while the dynamics of growth and reproduction phases can be unequally dependent on competition, as suggested here.
Competition in growing agricultural human communities is another example of such a situation. 
However, modification of the model is needed, if features of sexual reproduction --- as opposed to asexual reproduction of cells --- are critically important.

Following this work, suggestions for future research can be made.
On the experimental side, studies can be continued by making comparative measurements in concentrations of candidate growth factors, or by exposing each cell type to the fluid collected from the other type to measure growth variations. Additionally, monitoring temporal changes in cell-cycle phase distributions would also provide valuable data for testing the model’s predictions. On the modeling side, a natural complementary step is simulating the organoids using cell-based models, which can take spatial dependencies or possible roles of mechanical forces into account. Furthermore, alternative formulations of ligand uptake---such as receptor occupancy determining the G1-to-S transition~\cite{walker_integrated_2006}---could be explored to assess whether the observed dynamics is reproduced under different uptake assumptions.

\section{\label{sec:concluding_remarks}Concluding remarks}
In this study, we introduced a minimal mean-field model that 
proposes a possible  mechanism for cell competition. Our model 
incorporates cell cycle dynamics and unequal uptake capacities for secreted signaling ligands in cell cultures. This is inspired by biased competition observed between liver progenitor cells and colorectal cancer cells~\cite{Ana2024}. In this specific case, the growth stage corresponds to the cell cycle, and the shared, consumer-produced resource is the signaling ligands.
To maintain simplicity, we considered a minimal set of elements required to reproduce the experimental data. These include that competition takes place through a single biochemical factor, which is common to all cells due to rapid diffusion.
Our analysis revealed that if it is the case, three key elements are necessary to reproduce the experimentally observed outcome: (1) the secretion of the factor by cells in an autocrine/paracrine manner; (2) differences in the ability of cell types to respond to the factor; and (3) cell-cycle-dependent sensitivity to the factor. All three elements have been previously observed in the cell types studied here. We also showed that the strength of competitive interactions is governed by the disparity in the response capacities of the cell types.

The model presented here provides solid foundation for future research. On the modeling side, growth-stage-dependent competition can be further explored not only in the context of cell cultures, but also in sociophysical settings, such as ecological competition in expanding agricultural communities.

\section*{Data package}
The data package will be provided.

\begin{acknowledgments}
This work was supported by the Netherlands Organization for Scientific Research (NWO) through Start-Up Grant No. 740.018.013. The authors thank Maria Lamprou, Merel Elise van Luyk, Bryan Verhoef, Tyler Shendruk, and Michael Cates for valuable discussions.
\end{acknowledgments}

\bibliography{apssamp}

\onecolumngrid
\section*{Appendices}

\appendix
\renewcommand{\thefigure}{\Alph{section}\arabic{figure}}
\makeatletter
\@addtoreset{figure}{section}
\makeatother

\section{Derivation of solutions for pure populations}
In this section, we derive the analytical solutions to the dimensionless forms of Eqs.~2, 5 and 6 for pure wild-type and pure cancer organoids. 
For convenience, we present the  dimensionless form of Eqs.~1--6 of the manuscript:

\begin{equation}
\tilde{v}_{\text{G1}}^{(\text{W}, \text{C})}(\tilde{t})
\;=\;\tilde{\mu}_{(\text{W}, \text{C})}\,\tilde{\alpha}(\tilde{t}) = \tilde{\mu}_{(\text{W}, \text{C})}\,\tilde{M}(\tilde{t})/N(\tilde{t});
\label{eq:v_G1_W_C_nondim}
\end{equation}
\begin{equation}
    \frac{\partial \rho^{(\text{W,C})}}{\partial \tilde{t}} = -\frac{\partial}{\partial \phi}\left(\rho^{(\text{W,C})} \tilde{v}^{(\text{W,C})}\right)- \tilde{r}^{(\text{W,C})} \rho^{(\text{W,C})};
    \label{eq:general_continuity_nondim}
\end{equation}
\begin{equation}
    \tilde{v}_{\mathrm{G1}}^{(\mathrm{W}, \mathrm{C})}(\tilde{t})\,\rho^{(\mathrm{W}, \mathrm{C})}(\pi^-,\tilde{t})
    = \tilde{v}_{\mathrm{S}}\;\rho^{(\mathrm{W}, \mathrm{C})}(\pi^+,\tilde{t});
    \label{eq:BC_cont_nondim}
\end{equation}
\begin{equation}
\tilde{v}_{\mathrm{G1}}^{(\mathrm{W}, \mathrm{C})}(\tilde{t})\,\rho^{(\mathrm{W}, \mathrm{C})}(0,\tilde{t})
= 2\,\tilde{v}_{\mathrm{S}}\,\rho^{(\mathrm{W}, \mathrm{C})}(2\pi,\tilde{t});
\label{eq:BC_doubl_nondim}
\end{equation}
\begin{align}
\frac{d \tilde{M}}{d\tilde{t}}
&=\;\Bigl(\tilde{S} - \tilde{\mu}_{\text{W}} \,\tfrac{\tilde{M}}{N}\Bigr)\,W(\tilde{t})
\;+\;\Bigl(\tilde{S} - \tilde{\mu}_{\text{C}} \,\tfrac{\tilde{M}}{N}\Bigr)\,C(\tilde{t});
\label{eq:M_nondim}
\end{align}
\begin{equation}
    \frac{d W_\text{D}}{d\tilde{t}}
=\;\tilde{r}_\text{D} \int_{0}^{\pi} \rho^{\text{W}}(\phi, \tilde{t})\,d\phi.
\label{eq:W_D_nondim}
\end{equation}
From this point onward, $\rho_\mathrm{G1}$ denotes $\rho(\phi,\tilde{t})$ for $\phi \in (0,\pi)$, and $\rho_\mathrm{S}$ denotes $\rho(\phi,\tilde{t})$ for $\phi \in (\pi,2\pi)$. The letters W and C refer to wild-type and cancer, respectively.

\subsection*{Pure wild-type solution}
\label{sec:pure_wt_sol}
For pure organoids---comprising either wild-type or cancer cells---we assume that, on average, the organoids begin in a quasi-steady state and remain in that state throughout the experiments. In the quasi-steady state, the distribution of cells in $\phi$-space should not change, which means:
\begin{equation}
\frac{\partial}{\partial \tilde{t}} \left(\frac{1}{W(\tilde{t})}\,\rho^{\mathrm{W}}(\phi,\tilde{t})\right) = 0,
\label{eq:qss_cond_W}
\end{equation}
and
\begin{equation}
\frac{\partial}{\partial \tilde{t}} \left(\frac{W_\mathrm{D}(\tilde{t})}{W(\tilde{t})}\right) = 0.
\label{eq:qss_cond_W_D}
\end{equation}
Equation~\ref{eq:qss_cond_W} requires that the progression speed through the G1 phase is constant, which in turn, requires a constant equilibrium concentration of the growth factor. So, we have:  
\begin{equation}
\tilde{v}_{\mathrm{G1}}^{\mathrm{W}} = \tilde{\mu}_{\mathrm{W}} \tilde{\alpha}_{\mathrm{W}}.
\end{equation}
With this, Eqs.~\ref{eq:general_continuity_nondim},~\ref{eq:W_D_nondim}, and~\ref{eq:M_nondim} simplify to:
\begin{equation}
    \frac{\partial \rho_{\mathrm{G1}}^{\mathrm{W}}}{\partial \tilde{t}}
 \;+\; \tilde{\mu}_{\mathrm{W}} \tilde{\alpha}_{\mathrm{W}}
         \,\frac{\partial \rho_{\mathrm{G1}}^{\mathrm{W}}}{\partial \phi}
  \;+\; \tilde{r}_{\mathrm{D}}\,\rho_{\mathrm{G1}}^{\mathrm{W}}
 = 0;
 \label{eq:rho_g1_nondim_pure_W}
\end{equation}
\begin{equation}
    \frac{\partial \rho_{\mathrm{S}}^{\mathrm{W}}}{\partial \tilde{t}}
 \;+\; \tilde{v}_{\mathrm{S}}
         \,\frac{\partial \rho_{\mathrm{S}}^{\mathrm{W}}}{\partial \phi}
 = 0;
 \label{eq:rho_s_nondim_pure_W}
\end{equation}
\begin{equation}
    \frac{d W_\text{D}}{d\tilde{t}}
=\;\tilde{r}_\text{D}\,W_{\text{G1}}(\tilde{t});
\end{equation}
\begin{equation}
\frac{d\,\tilde{M}}{d\tilde{t}}
 \;=\;\biggl(\tilde{S} \;-\; \tilde{\mu}_{\mathrm{W}}
        \tilde{\alpha}_{\mathrm{W}}\biggr)
        \,W(\tilde{t})
  \;=\; \tilde{S} W(\tilde{t}) \;-\; \tilde{\mu}_{\mathrm{W}}\tilde{M}(\tilde{t}).
  \label{eq:M_tilde_pure_W}
\end{equation}
According to the experiments for the pure wild-type organoids: 
\begin{equation}
    W(\tilde{t}) = W(0)\;e^{\tilde{t}},
    \label{eq:W_exp_pure_W}
\end{equation}
and the differentiating cells:
\begin{equation}
    W_\mathrm{D}(\tilde{t}) = f_\mathrm{D} W(\tilde{t}) = f_\mathrm{D}\;W(0)\;e^{\tilde{t}},
\end{equation}
where \(f_\mathrm{D}\) is the fraction of differentiating wild-type cells, which approximately equals to 0.2 according to the experiments reported in Ref.~\citenum{Ana2024}.
Given that \(\tilde{M}(\tilde{t})=\tilde{\alpha}_{\mathrm{W}} W(\tilde{t})\) and by substituting Eq.~\ref{eq:W_exp_pure_W} in Eq.~\ref{eq:M_tilde_pure_W}, we find the equilibrium concentration for the pure wild-type organoids:
\begin{equation}
    \tilde{\alpha}_{\mathrm{W}} = {{\tilde{S}}\over {1+\tilde{\mu}_{\mathrm{W}}}}.
    \label{eq:alpha_w}
\end{equation}
We use this equilibrium value for the initial condition of \(\tilde{M}\):
\begin{equation}
\tilde{M}(0) = \tilde{\alpha}_{\mathrm{W}}\, W(0).
\label{eq:init_cond_M_pure_W}
\end{equation}
Now we find the density of cells in cell phase. Combining Eqs.~\ref{eq:qss_cond_W} and~\ref{eq:W_exp_pure_W} results in:
\begin{align}
\rho^{\mathrm{W}}_{\mathrm{G1}}(\phi,\tilde{t}) 
&= f_{\mathrm{W}}(\phi)\,e^{\tilde{t}} 
\quad \text{for} \quad \phi \in (0,\pi); \notag\\
\rho^{\mathrm{W}}_{\mathrm{S}}(\phi,\tilde{t}) 
&= h_{\mathrm{W}}(\phi)\,e^{\tilde{t}} 
\quad \text{for} \quad \phi \in (\pi, 2\pi).
\end{align}
By substituting these equations into Eqs.~\ref{eq:rho_g1_nondim_pure_W} and ~\ref{eq:rho_s_nondim_pure_W}, the functions \(f_{\mathrm{W}}(\phi)\) and \(h_{\mathrm{W}}(\phi)\) are found (up to a constant factor) as

\begin{equation}
f_{\mathrm{W}}(\phi) = f_{\mathrm{W}}(0)\,\exp\!\left(-\frac{1 + \tilde{r}_{\mathrm{D}}}{\tilde{\mu}_{\mathrm{W}} \tilde{\alpha}_{\mathrm{W}}}\,\phi\right);
\end{equation}
\begin{equation}
h_{\mathrm{W}}(\phi) = h_{\mathrm{W}}(\pi)\,\exp\!\left(-\frac{\phi - \pi}{\tilde{v}_{\mathrm{S}}}\right).
\end{equation}
These functions show exponential distributions for the density of cells in \(\phi\). Consequently, we will have the densities:

\begin{equation}
    \rho^{\mathrm{W}}_{\mathrm{G1}}(\phi,\tilde{t}) = \rho^{\mathrm{W}}_{\mathrm{G1}}(0,0)\,\exp\!\left(-\frac{1 + \tilde{r}_{\mathrm{D}}}{\tilde{\mu}_{\mathrm{W}} \tilde{\alpha}_{\mathrm{W}}}\,\phi\right) \,e^{\tilde{t}};
\end{equation}

\begin{equation}
\rho^{\mathrm{W}}_{\mathrm{S}}(\phi,\tilde{t}) = \rho^{\mathrm{W}}_{\mathrm{S}}(\pi,0) \,\exp\!\left(-\frac{\phi - \pi}{\tilde{v}_{\mathrm{S}}}\right) \,e^{\tilde{t}}.
\end{equation}
By applying the boundary conditions in Eqs.~\ref{eq:BC_cont_nondim} and~\ref{eq:BC_doubl_nondim}, we have:
\begin{equation}
    \rho^{\mathrm{W}}_{\mathrm{S}}(\pi,0) = \rho^{\mathrm{W}}_{\mathrm{G1}}(0,0)\,{{\tilde{\mu}_{\mathrm{W}} \tilde{\alpha}_{\mathrm{W}}}\over{\tilde{v}_{\mathrm{S}}}} \,e^{-m\pi},
\end{equation}
and
\begin{equation}
    m + {1 \over \tilde{v}_{\mathrm{S}}} = {{\ln{2}}\over{\pi}},
    \label{eq:BC_div_with_m}
\end{equation}
where for brevity, we  {have introduced} \(m\):
\begin{equation}
m = \frac{1 + \tilde{r}_{\mathrm{D}}}{\tilde{\mu}_{\mathrm{W}} \tilde{\alpha}_{\mathrm{W}}} = \frac{1 + \tilde{r}_{\mathrm{D}}}{\tilde{\mu}_{\mathrm{W}} {{{\tilde{S}}\over {1+\tilde{\mu}_{\mathrm{W}}}}}}.
\label{eq:m_expression}
\end{equation}
The factor \(\rho^{\mathrm{W}}_{\mathrm{G1}}(0,0)\) determines the initial population of wild-type cells, because:
\begin{equation}
    \int^{2 \pi}_{0} \rho^{\mathrm{W}}(\phi,0)  \;d\phi = (1-f_\mathrm{D})W(0).
\end{equation}
Hence, the expression for \(\rho^{\mathrm{W}}_{\mathrm{G1}}(0,0)\) is:
\begin{equation}
    \begin{split}
    \rho^{\mathrm{W}}_{\mathrm{G1}}(0,0) = &\; (1-f_\mathrm{D})W(0)\bigg(\frac{1}{m}(1-e^{-m\pi}) +\tilde{\mu}_\mathrm{W}\,\tilde{\alpha}_\mathrm{W}\,e^{-m\pi}\,(1-e^{-\pi/\tilde{v}_\mathrm{S}})\bigg)^{-1}\\
    = &\; 2m(1-f_\mathrm{D})W(0)/\left(1+\tilde{r}_\mathrm{D}\left(e^{\pi/\tilde{v}_\mathrm{S}}-1\right)\right).
    \end{split}
\end{equation}
Additionally, we force to keep the fraction of differentiating cells equal to \(f_{\mathrm{D}} = 0.2\), which is based on the experimental data. Therefore, we find
\begin{equation}
    \frac{d W_{\mathrm{D}}}{d\tilde{t}} = f_\mathrm{D}\;W(0)\;e^{\tilde{t}}=\tilde{r}_\mathrm{D}W_{\mathrm{G1}}(\tilde{t})=\tilde{r}_\mathrm{D} \int^{\pi}_{0} \rho^{\mathrm{W}}_{\mathrm{G1}}(\phi,\tilde{t})  \;d\phi,
\end{equation}
which leads to
\begin{equation}
    (1+\tilde{r}_\mathrm{D})\bigg( \frac{1-e^{-\pi/\tilde{v}_\mathrm{S}}}{e^{m \pi}-1}\bigg)+1-\tilde{r}_\mathrm{D}\bigg( \frac{1-f_\mathrm{D}}{f_\mathrm{D}}\bigg) = 0,
    \label{eq:f_D_constant_condition}
\end{equation}
where the value of \(m\) is as introduced in Eq.~\ref{eq:m_expression}. 
Now, by substituting $m$ from Eq.~\ref{eq:BC_div_with_m} into Eq.~\ref{eq:f_D_constant_condition}, we obtain:
\begin{equation}
    \tilde{r}_\mathrm{D} = f_\mathrm{D}/\left({2-f_\mathrm{D} - \exp(\pi/\tilde{v}_\mathrm{S})}\right).
\end{equation}
Consequently, combining Eqs.~\ref{eq:BC_div_with_m} and~\ref{eq:m_expression}, leads to:
\begin{equation}
\tilde{\mu}_\mathrm{W} = \left(\frac{\tilde{S}}{1+\tilde{r}_\mathrm{D}}\left(\frac{\ln2}{\pi} - \frac{1}{\tilde{v}_\mathrm{S}}\right) -1\right)^{-1} = 1/\left( \tilde{S} m/\left({1+\tilde{r}_\mathrm{D}}\right) -1\right).
\end{equation}
In summary, the quasi-steady state solution for the pure wild-type cells is:  
\begin{tcolorbox}[colback=white, colframe=black, title={Pure wild-type solution}]
\begin{equation}
    \rho^{\mathrm{W}}_{\mathrm{G1}}(\phi,\tilde{t}) = \rho^{\mathrm{W}}_{\mathrm{G1}}(0,0)\,\exp\!\left(-\frac{1 + \tilde{r}_{\mathrm{D}}}{\tilde{\mu}_{\mathrm{W}} \tilde{\alpha}_{\mathrm{W}}}\,\phi\right) \,e^{\tilde{t}};
    \label{eq:rho_g1_pure_W_sol}
\end{equation}
\begin{equation}
    \rho^{\mathrm{W}}_{\mathrm{S}}(\phi,\tilde{t}) = \rho^{\mathrm{W}}_{\mathrm{G1}}(0,0)\,e^{-m\pi}\,{{\tilde{\mu}_{\mathrm{W}} \tilde{\alpha}_{\mathrm{W}}}\over{\tilde{v}_{\mathrm{S}}}} \,\exp\!\left(-\frac{\phi - \pi}{\tilde{v}_{\mathrm{S}}}\right) \,e^{\tilde{t}};
    \label{eq:rho_s_pure_W_sol}
\end{equation}
\begin{equation}
    W_\mathrm{D}(\tilde{t}) =  f_\mathrm{D}\;W(0)\;e^{\tilde{t}};
    \label{eq:W_D_pure_W_sol_app}
\end{equation}
\begin{equation}
    \tilde{M}(\tilde{t}) = \tilde{\alpha}_\mathrm{W} \;W(0)\;e^{\tilde{t}}.
    \label{eq:M_pure_W_sol_app}
\end{equation}
\end{tcolorbox}

\subsection*{Pure cancer solution}
\label{sec:pure_c_sol}
For pure cancer organoids, we adopt an approach analogous to that used for pure wild-type population. Under quasi-steady-state conditions:
\begin{equation}
\frac{\partial}{\partial \tilde{t}} \left(\frac{1}{C(\tilde{t})}\,\rho^{\mathrm{C}}(\phi,\tilde{t})\right) = 0.
\label{eq:qss_cond_C}
\end{equation}
Thus, the progression speed of cells in the G1 phase, and the concentration of the growth factor remain constant. So, we have:
\begin{equation}
\tilde{v}_{\mathrm{G1}}^{\mathrm{C}} = \tilde{\mu}_{\mathrm{C}} \tilde{\alpha}_{\mathrm{C}},
\end{equation}
where \(\tilde{\alpha}_{\mathrm{C}}\) is the equilibrium concentration of growth factor in pure cancer organoids.
Therefore, the Eqs.~\ref{eq:general_continuity_nondim} and~\ref{eq:M_nondim} read:
\begin{equation}
    \frac{\partial \rho_{\mathrm{G1}}^{\mathrm{C}}}{\partial \tilde{t}}
 \;+\; \tilde{\mu}_{\mathrm{C}} \tilde{\alpha}_{\mathrm{C}}
         \,\frac{\partial \rho_{\mathrm{G1}}^{\mathrm{C}}}{\partial \phi}
 = 0;
 \label{eq:rho_g1_nondim_pure_C}
\end{equation}
\begin{equation}
    \frac{\partial \rho_{\mathrm{S}}^{\mathrm{C}}}{\partial \tilde{t}}
 \;+\; \tilde{v}_{\mathrm{S}}
         \,\frac{\partial \rho_{\mathrm{S}}^{\mathrm{C}}}{\partial \phi}
 = 0;
 \label{eq:rho_s_nondim_pure_C}
\end{equation}
\begin{equation}
\frac{d\,\tilde{M}}{d\tilde{t}}
 \;=\;\biggl(\tilde{S} \;-\; \tilde{\mu}_{\mathrm{C}}
        \tilde{\alpha}_{\mathrm{C}}\biggr)
        \,C(\tilde{t})
  \;=\; \tilde{S} C(\tilde{t}) \;-\; \tilde{\mu}_{\mathrm{C}}\tilde{M}(\tilde{t}).
  \label{eq:M_tilde_pure_C}
\end{equation}
From the experiments:
\begin{equation}
    C(\tilde{t}) = C(0)\;e^{\beta_\mathrm{C}\frac{\tilde{t}}{\beta_\mathrm{W}}} = C(0)\;e^{\gamma \tilde{t}},
    \label{eq:C_exp_pure_C}
\end{equation}
where we denote the ratio \(\beta_\mathrm{C}/\beta_\mathrm{W}\) by \(\gamma\). By substituting \(\tilde{M}(\tilde{t})=\tilde{\alpha}_{\mathrm{C}} C(\tilde{t})\) and Eq.~\ref{eq:C_exp_pure_C} in Eq.~\ref{eq:M_tilde_pure_C}, we can find \(\tilde{\alpha}_\mathrm{C}\):
\begin{equation}
    \tilde{\alpha}_\mathrm{C} = \frac{\tilde{S}}{\gamma+\tilde{\mu}_\mathrm{C}}.
\end{equation}
For the initial condition:
\begin{equation}
\tilde{M}(0) = \tilde{\alpha}_{\mathrm{C}}\, C(0).
\label{eq:init_cond_M_pure_C}
\end{equation}
Combining Eq.~\ref{eq:qss_cond_C} with Eq.~\ref{eq:C_exp_pure_C} results in:
\begin{equation}
\rho^{\mathrm{C}}_{\mathrm{G1}}(\phi,\tilde{t}) = f_{\mathrm{C}}(\phi)\,e^{\gamma \tilde{t}} 
\quad \textrm{for} \quad
\phi \in (0,\pi)\; ;
\end{equation}
\begin{equation}
\rho^{\mathrm{C}}_{\mathrm{S}}(\phi,\tilde{t}) = h_{\mathrm{C}}(\phi)\,e^{\gamma \tilde{t}}, 
\quad \textrm{for} \quad
\phi \in (\pi, 2\pi).
\end{equation}
By substituting these equations into Eqs.~\ref{eq:rho_g1_nondim_pure_C} and~\ref{eq:rho_s_nondim_pure_C} we have the following.
\begin{equation}
f_{\mathrm{C}}(\phi) = f_{\mathrm{C}}(0)\,\exp\!\left(-\frac{\gamma}{\tilde{\mu}_{\mathrm{C}} \tilde{\alpha}_{\mathrm{C}}}\,\phi\right);
\end{equation}
\begin{equation}
h_{\mathrm{C}}(\phi) = h_{\mathrm{C}}(\pi)\,\exp\!\left(- \frac{\gamma}{\tilde{v}_{\mathrm{S}}} (\phi - \pi)\right).
\end{equation}
Hence, for the densities are found:
\begin{equation}
    \rho^{\mathrm{C}}_{\mathrm{G1}}(\phi,\tilde{t}) = \rho^{\mathrm{C}}_{\mathrm{G1}}(0,0)\,\exp\!\left(-\frac{\gamma}{\tilde{\mu}_{\mathrm{C}} \tilde{\alpha}_{\mathrm{C}}}\,\phi\right) \,e^{\gamma \tilde{t}}; 
\end{equation}
\begin{equation}
\rho^{\mathrm{C}}_{\mathrm{S}}(\phi,\tilde{t}) = \rho^{\mathrm{C}}_{\mathrm{S}}(\pi,0) \,\exp\!\left(- \frac{\gamma}{\tilde{v}_{\mathrm{S}}} (\phi - \pi)\right) \,e^{\gamma \tilde{t}}.
\end{equation}
From the boundary conditions for cancer cells in Eqs.~\ref{eq:BC_cont_nondim} and~\ref{eq:BC_doubl_nondim}, we derive:
\begin{equation}
    \rho^{\mathrm{C}}_{\mathrm{S}}(\pi,0) = \rho^{\mathrm{C}}_{\mathrm{G1}}(0,0)\,e^{-q\pi}\,{{\tilde{\mu}_{\mathrm{C}} \tilde{\alpha}_{\mathrm{C}}}\over{\tilde{v}_{\mathrm{S}}}},
\end{equation}
and
\begin{equation}
    q + {\gamma \over \tilde{v}_{\mathrm{S}}} = {{\ln{2}}\over{\pi}},
    \label{eq:BC_div_with_q}
\end{equation}
where \(q\) is:
\begin{equation}
q = \frac{\gamma}{\tilde{\mu}_{\mathrm{C}} \tilde{\alpha}_{\mathrm{C}}} = \frac{\gamma}{\tilde{\mu}_{\mathrm{C}} {{{\tilde{S}}\over {\gamma+\tilde{\mu}_{\mathrm{C}}}}}}.
\label{eq:q_expression}
\end{equation}
The factor \(\rho^{\mathrm{C}}_{\mathrm{G1}}(0,0)\) is determined by the initial number of cells:
\begin{equation}
    \int^{2 \pi}_{0} \rho^{\mathrm{C}}(\phi,0)  \;d\phi = C(0), 
\end{equation}
so we find:
\begin{equation}
\begin{split}
\rho^{\mathrm{C}}_{\mathrm{G1}}(0,0) 
&= C(0)\bigg(
    \frac{1}{q}(1 - e^{-q\pi})
    + \frac{\tilde{\mu}_\mathrm{C}\,\tilde{\alpha}_\mathrm{C}}{\gamma}
      e^{-q\pi}(1 - e^{-\gamma \pi / \tilde{v}_\mathrm{S}})
    \bigg)^{-1} \\
&= 2 q\, C(0).
\end{split}
\end{equation}
and
\begin{equation}
    \rho_\mathrm{S}^{\mathrm{C}}(\pi^+,0)
    =  C(0)\,\left({\gamma/{\tilde{v}_\mathrm{S}}}\right) \,e^{\gamma\pi/\tilde{v}_\mathrm{S}}.
\end{equation}
Finally, combining Eqs.~\ref{eq:q_expression} and~\ref{eq:BC_div_with_q} leads to:
\begin{equation}
    \tilde{\mu}_\mathrm{C} =\gamma \left(\frac{\tilde{S}}{\gamma}\left(\frac{\ln2}{\pi} - \frac{\gamma}{\tilde{v}_\mathrm{S}}\right) -1\right)^{-1} = \gamma^2/ \left(\tilde{S}q-\gamma\right).
\end{equation}
Hence, the complete solution for pure cancer organoids is as follows.  
\begin{tcolorbox}[colback=white, colframe=black, title={Pure cancer solution}]
\begin{equation}
    \rho^{\mathrm{C}}_{\mathrm{G1}}(\phi,\tilde{t}) = \rho^{\mathrm{C}}_{\mathrm{G1}}(0,0)\,\exp\!\left(-\frac{\gamma}{\tilde{\mu}_{\mathrm{C}} \tilde{\alpha}_{\mathrm{C}}}\,\phi\right) \,e^{\gamma \tilde{t}};
    \label{eq:rho_g1_pure_C_sol}
\end{equation}
\begin{equation}
    \rho^{\mathrm{C}}_{\mathrm{S}}(\phi,\tilde{t}) = \rho^{\mathrm{C}}_{\mathrm{G1}}(0,0)\,e^{-q\pi}\,{{\tilde{\mu}_{\mathrm{C}} \tilde{\alpha}_{\mathrm{C}}}\over{\tilde{v}_{\mathrm{S}}}} \,\exp\!\left(- \frac{\gamma}{\tilde{v}_{\mathrm{S}}} (\phi - \pi)\right) \,e^{\gamma \tilde{t}};
    \label{eq:rho_s_pure_C_sol}
\end{equation}
\begin{equation}
    \tilde{M}(\tilde{t}) = \tilde{\alpha}_\mathrm{C} \;C(0)\;e^{\gamma \tilde{t}}.
    \label{eq:M_pure_C_sol_app}
\end{equation}
\end{tcolorbox}

\newpage
\onecolumngrid
\section{The distributions of the initial number of cells}
\label{app:init_dist}

\begin{figure}[h]
\centering
\includegraphics[width=1.0\textwidth]{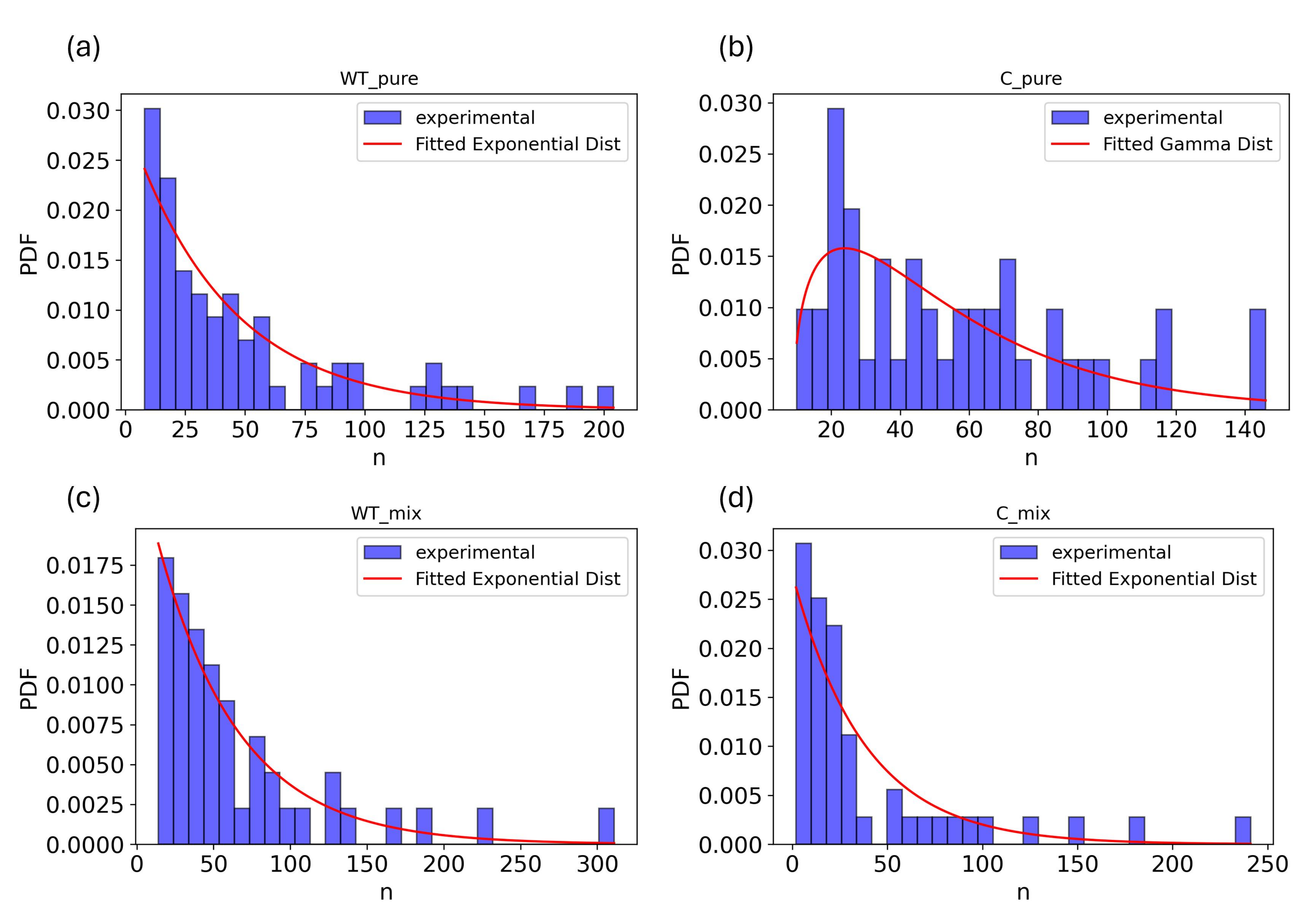}
\caption{The experimental data for the initial number of cells in organoids, as well as the fitted distributions. The panels show the data for (a) pure wild-type cells,  (b) pure cancer cells,  (c) wild-type cells in mixed  conditions and (d) cancer cells in mixed  {conditions.} The experimental data was taken from Ref.~\citenum{Ana2024}. The red curves indicate the fitted distribution to each set of data.}
\label{fig:init_dist}
\end{figure}

As a input of the model, we needed samples for initial number of cells in mixed organoids. In order to obtain these statistics, we used the initial numbers of cells in the experiments. We first fitted distributions to the experimental data on the initial number of cells. These fittings as well as the data (taken from Ref.~\citenum{Ana2024}) are shown in Fig.~\ref{fig:init_dist}. 
For cancer cells in their pure conditions we used Gamma distribution, shown in Eq.~\ref{eq:gamma_dist}, while for the other three cases we used exponential distribution, shown in Eq.~\ref{eq:expo_dist}.
\begin{equation}
    f(n) = \frac{(n-n_0)^{\alpha-1}}{\Gamma(\alpha) \;\theta^{\alpha}}\;\exp{\left(-\frac{n-n_0}{\theta} \right)}\quad \text{for} \quad n>n_0;
    \label{eq:gamma_dist} 
\end{equation}
\begin{equation}
    f(n) = \frac{1}{\theta}\;\exp{\left(-\frac{n-n_0}{\theta} \right)} \quad \text{for} \quad n>n_0.
    \label{eq:expo_dist}
\end{equation}
For the fittings, we used SciPy package in Python, that uses maximum likelihood estimation (MLE) to find the parameters of the fitted functions. The parameters of the fitted functions are presented in Table~\ref{tab:init_fit_params}. Once we found the fitted distributions, we drew random samples with these distributions to use in the initial condition of the model.
It is worth noting that for pure organoids, the initial number of cells was always set to one. Hence, only the distributions shown in panels~(c) and~(d) of 
Fig.~\ref{fig:init_dist} were used.
The fittings for pure organoids are included for completeness.

\begin{table}\centering
\caption{Parameters of the fitted distributions to the initial number of cells in pure and mixed organoids.
The parameters \(\alpha\), \(\theta\) and \(n_0\) {are the shape, scale, and shift parameters, respectively.}
}
\begin{tabular}{lrrr}
  & $\alpha$ & $n_0$ & $\theta$ \\
\midrule
pure wild-type & - & 8.0 & 41.47 \\
pure cancer & 1.47 & 9.12 & 31.37 \\
mixed wild-type & - & 14.0 & 53.04 \\
mixed cancer & - & 2.0 & 38.18\\
\bottomrule
\end{tabular}
\label{tab:init_fit_params}
\end{table}

\newpage
\section{Cost function}
\label{app:cost_func}

As mentioned in the main text, the cost function {comprises} three terms. The first term is related to the populations of wild-type and cancer cells in mixed organoids, the second term is for reproduction of the statistics of wild-type cells in different phases, and the third term for the effect of initial composition of organoids on the growth of each population. The total cost is calculated as in Eq.~\ref{eq:cost_function_append}, which is the same as Eq.~28 in the manuscript.
\begin{equation}
    L(\tilde{S}, \tilde{v}_\mathrm{S}) = 
    \frac{1}{3}
    L_{\text{pop}}(\tilde{S}, \tilde{v}_\mathrm{S}) + 
    \frac{1}{3}
    L_{\text{stat}}(\tilde{S}, \tilde{v}_\mathrm{S}) + 
    \frac{1}{3}
    L_{\text{init}}(\tilde{S}, \tilde{v}_\mathrm{S}).
    \label{eq:cost_function_append}
\end{equation}

The first term, \(L_{\text{pop}}\) is calculated using \(W^*(t) = W(t)/W(0)\) and  \(C^*(t) = C(t)/C(0)\), the normalized populations of wild-type and cancer cells. Since the baseline for the population dynamics follows exponential growth, it is more appropriate to use the logarithms of
\(W^*(t)\) and \(C^*(t)\) 
in the definition of the error. Thus, we have:
\begin{equation}
    L_{\text{pop}}=
    \frac{1}{2}\sum_i \omega_i^\text{W} \left[ \ln{\left(W^*_{\text{model}}(t_i)\right)}- \ln{\left(W^*_{\text{exp}}(t_i)\right)}\right]^2
    +
    \frac{1}{2}\sum_i \omega_i^\text{C} \left[ \ln{\left(C^*_{\text{model}}(t_i)\right)}- \ln{\left(C^*_{\text{exp}}(t_i)\right)}\right]^2
\end{equation}
where \(i\) counts on the experimental data points, and the weight factors, \(\omega_i^\text{W}\) and \(\omega_i^\text{C}\) are calculated using the uncertainties for the experimental data, \(\sigma_{\text{exp}}^\text{W}\) and \(\sigma_{\text{exp}}^\text{C}\):
\begin{equation}
    \omega_i^\text{W} = \frac{1}{{|\sigma_{\text{exp}}^\text{W} (t_i)}/W_{\text{exp}}^{*}(t_i)|^2}
    \left[ 
    \sum_j \frac{1}{{|\sigma_{\text{exp}}^\text{W} (t_j)}/W_{\text{exp}}^{*}(t_j)|^2}
    \right]^{-1},
\end{equation}
and
\begin{equation}
    \omega_i^\text{C} = \frac{1}{{|\sigma_{\text{exp}}^\text{C} (t_i)}/C_{\text{exp}}^{*}(t_i)|^2}
    \left[ 
    \sum_j \frac{1}{{|\sigma_{\text{exp}}^\text{C} (t_j)}/C_{\text{exp}}^{*}(t_j)|^2}
    \right]^{-1}.
\end{equation}
In order to make different terms of cost function to have the same order of magnitude in their numerical values, we need to multiply error terms by factors that are calculated based on typical scales of each of each cost terms. For the scale of the first term, we consider:
\begin{equation}
    L_{\text{pop}}^0 = 
    \frac{1}{2}
    \left\langle 
    \left(    \frac{\sigma_{\text{exp}}^{\text{W}}(t_i)}{W_{\text{exp}}^{*}(t_i)}
    \right)^2
    \right\rangle_i
    +
    \frac{1}{2}
    \left\langle 
    \left(    \frac{\sigma_{\text{exp}}^{\text{C}}(t_i)}{C_{\text{exp}}^{*}(t_i)}
    \right)^2
    \right\rangle_i = 4.9\times10^{-3},
\end{equation}
where the averaging is done over the experimental data points.

For the second term, the error of statistics of wild-type cells in different phases, point out several requirements. Ideally, we want to fix the percentages of wild-type in three different states, for pure and mixed organoids. This means 6 quantities must be fixed. Since for each case (either pure or mixed), the fractions must sum up to 1.0, we need to fix 4 independent quantities. However, the fraction of differentiated wild-type cells (\(f_\text{D}\)) in pure  {conditions} is already fixed in the model by Eq.~\ref{eq:f_D_constant_condition}. Hence, we need to fix three quantities. We choose the first one to be fraction of wild-type cells in S/G2/M phases in pure organoids.  The second quantity is the ratio of the fractions of wild-type cells in mixed and pure organoids, \(R_{\text{S}} =\frac{f_\text{S,W,mix}}{f_\text{S,W,pure}}\). The third one is the same ratio, but for wild-type cells in G0 phase, i.e., \(R_{\text{G0}} = \frac{f_\text{G0,W,mix}}{f_\text{G0,W,pure}}\). From the experimental data, the target values for these ratios are \( R_{\text{S}}^{\text{tar}} = 0.5(\frac{30}{47}+\frac{10}{20}) =0.57\) and \(R_{\text{G0}}^{\text{tar}} = \frac{29}{20}=1.45\). We have used the experimental data presented in Fig.~4 in the manuscript to calculate these target values. Consequently, we consider the cost function of the statistics as follows
\begin{equation}
    L_{\text{stat}} =
    \left(
    \frac{1}{3}
    L_{\text{S,pure}}
    +
    \frac{1}{3}
    L_{\text{S,mix}}
    +
    \frac{1}{3}
    L_{\text{G0,mix}}
    \right)
    \frac{L^0_{\text{pop}}}{L^0_{\text{stat}}},
\end{equation}
where
\begin{equation}
    L_{\text{S,pure}} = \left(f_{\text{S,W,pure}}^{\text{model}}-f_{\text{S,W,pure}}^{\text{exp}}\right)^2;
\end{equation}
\begin{equation}
    L_{\text{S,mix}} = \left(f_{\text{S,W,pure}}^{\text{model}}-{R_{\text{S}}^{\text{tar}}}\;f_{\text{S,W,pure}}^{\text{model}}\right)^2;
\end{equation}
\begin{equation}
    L_{\text{G0,mix}} = \left(f_{\text{G0,W,pure}}^{\text{model}}-{R_{\text{G0}}^{\text{tar}}}\;f_{\text{G0,W,pure}}^{\text{model}}\right)^2.
\end{equation}
The factor \( L^0_{\text{pop}}/L^0_{\text{stat}}\) makes this error term numerically comparable to the first term \(L_{\text{pop}}\).
According to the experimental data, the uncertainty for the fractions of wild-type cells in the G0 and S/G2/M phases is approximately \(2.0\sim3.0\%\). So, we consider the scale of this term as 
\begin{equation}
    L_{\text{stat}}^{0} = 0.03^2=9.0\times10^{-4}.
\end{equation}

For the last term of cost function, \(L_{\text{init}}\), we take the following path. First, we fit lines to the logarithm of experimental growth of each type at \(t=60~\text{h}\) --\(\ln{(W^*_{\text{exp}}(60))}\) or \(\ln{(C^*_{\text{exp}}(60))}\)-- as a function of the initial percentage of each cell type, i.e., \(F_{\text{W}}(0)\) and \(F_{\text{C}}(0)\) for wild-type and cancer cells, respectively. The data are the same as shown in Fig.~3 in the manuscript. We assumed that for wild-type cells the fitted line is \(y=A+Bx\) and for cancer cells it is \(y=D+Ex\). Least squares fitting provided us with the coefficients \(A=0.749\), \(B=0.570\), \(D=3.362\) and \(E=-1.966\). This allows us to write the third cost-function term as:
\begin{equation}
    L_{\text{init}} = 
    \frac{1}{2}
    \frac{L_{\text{pop}}^0}{L_{\text{init}}^{\text{W,0}}}
    L_{\text{init}}^{\text{W}}
    +
    \frac{1}{2}
    \frac{L_{\text{pop}}^0}{L_{\text{init}}^{\text{C,0}}}
    L_{\text{init}}^{\text{C}},
\end{equation}
where the error terms are
\begin{equation}
    {L_{\text{init}}^{\text{W}}} = 
    \left\langle
    \left[\;
    \ln \left(W^*_{\text{model}}(60)\right) - \left(A+B F_{\text{W}}(0)\right)
    \;\right]^2
    \right\rangle_{\text{model}} ,
\end{equation}
and
\begin{equation}
    {L_{\text{init}}^{\text{C}}} = 
    \left\langle
    \left[\;
    \ln {\left(C^*_{\text{model}}(60)\right)} - \left(D+E F_{\text{C}}(0)\right)
    \;\right]^2
    \right\rangle_{\text{model}}.
\end{equation}
Again, the factors 
\({L_{\text{pop}}^0}/{L_{\text{init}}^{\text{W,0}}}\)
and 
\({L_{\text{pop}}^0}/{L_{\text{init}}^{\text{C,0}}}\) 
rescale the error terms to be numerically comparable to \(L_{\text{pop}}\),
and the scales 
\(L_{\text{init}}^{\text{W,0}}\) and 
\(L_{\text{init}}^{\text{C,0}}\) 
are calculated as
\begin{equation}
    {L_{\text{init}}^{\text{W,0}}} = 
    \mathrm{Var}\left[ 
    \ln {\left(W^*_{\text{exp}}(60)\right)} - \left(A+B F_{\text{W}}(0)\right)
    \right]_{\text{exp}}
    =
    0.094;
\end{equation}
\begin{equation}
    {L_{\text{init}}^{\text{C,0}}} = 
    \mathrm{Var}\left[ 
    \ln {\left(C^*_{\text{exp}}(60)\right)} - \left(D+E F_{\text{C}}(0)\right)
    \right]_{\text{exp}}
    =
    0.26.
\end{equation}

\newpage
\section{Temporal variations of phases}
\label{app:temporal}

\begin{figure}[h]
\centering
\includegraphics[width=1.0\textwidth]{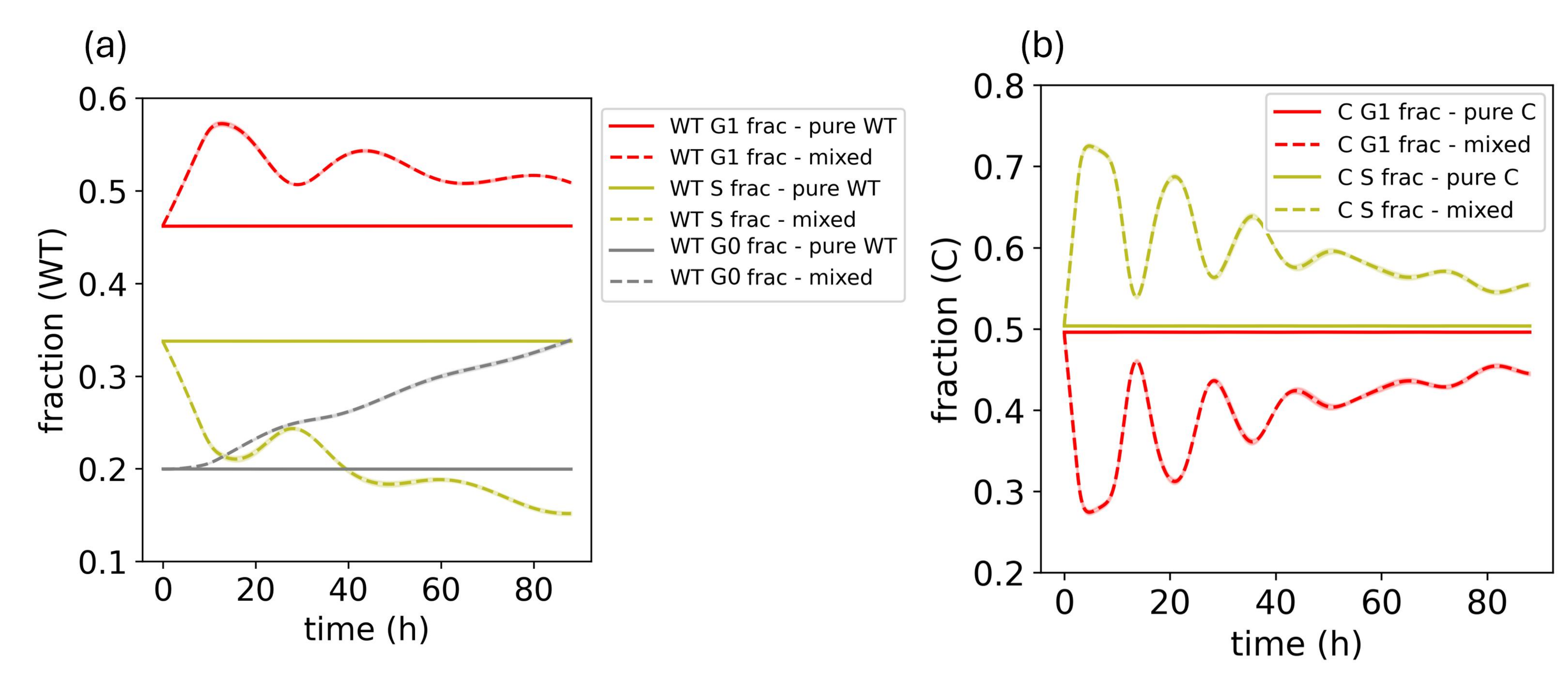}
\caption{Evolution of cell-cycle phase fractions over time. 
(a) Fractions of wild-type cells in the G1, S/G2/M, and G0 (differentiated) phases as a function of time. 
(b) Fractions of cancer cells in the G1 and S/G2/M phases over time. 
In both panels, solid lines represent pure organoids, while dashed lines indicate mixed conditions. 
Colors denote specific phases: red for G1, yellow for S/G2/M, and gray for G0 state. 
Note that cancer cells do not enter the differentiated phase. }
\label{fig:temporal_fractions}
\end{figure}

As a complementary analysis, we present the model’s predictions for the temporal evolution 
of phase-specific cel.l fractions in Fig.~\ref{fig:temporal_fractions}. In  pure  
wild-type and cancer organoids, these fractions remain constant over time—a natural consequence of 
the quasi-steady-state assumption for the normalized cell density distribution.
In mixed organoids, the growth factor concentration lies between the equilibrium values 
of pure wild-type and pure cancer cells.
According to 
Eq.~1 in the manuscript, this results in a lower G1 progression 
speed for wild-type cells and a higher speed for cancer cells, compared to their respective 
pure cases.
These differences lead to a reduction in the G1-to-S flux for wild-type cells, 
and an increase for cancer cells. Given that the S/G2/M progression speed \(v_\textrm{S}\) 
is assumed to be constant, wild-type cells accumulate in the G1 phase, while cancer cells become depleted. Consequently, the opposite occurs in the S/G2/M phases. In addition, due to the increased G1 fraction in wild-type cells, the fraction in G0 increases monotonically, as described by Eq.~6 in the main text. Moreover, these dynamics give rise to temporal fluctuations 
in the phase fractions, as the accumulation and depletion waves induced by mixing, propagate through 
the cell cycle and repeat periodically. However, these fluctuations gradually diminish as the total 
population grows, and the concentration asymptotically approaches the equilibrium value of cancer cells. Experimental data on the fractions of cancer cells, as well as the temporal variations for both types, are still lacking. Future measurements in this direction would allow for a more rigorous evaluation of the assumptions and predictions of the model.

\newpage
\section{Sensitivity analysis}
\label{app:sensitivity}
\begin{figure}
\centering
\includegraphics[width=1.0\textwidth]{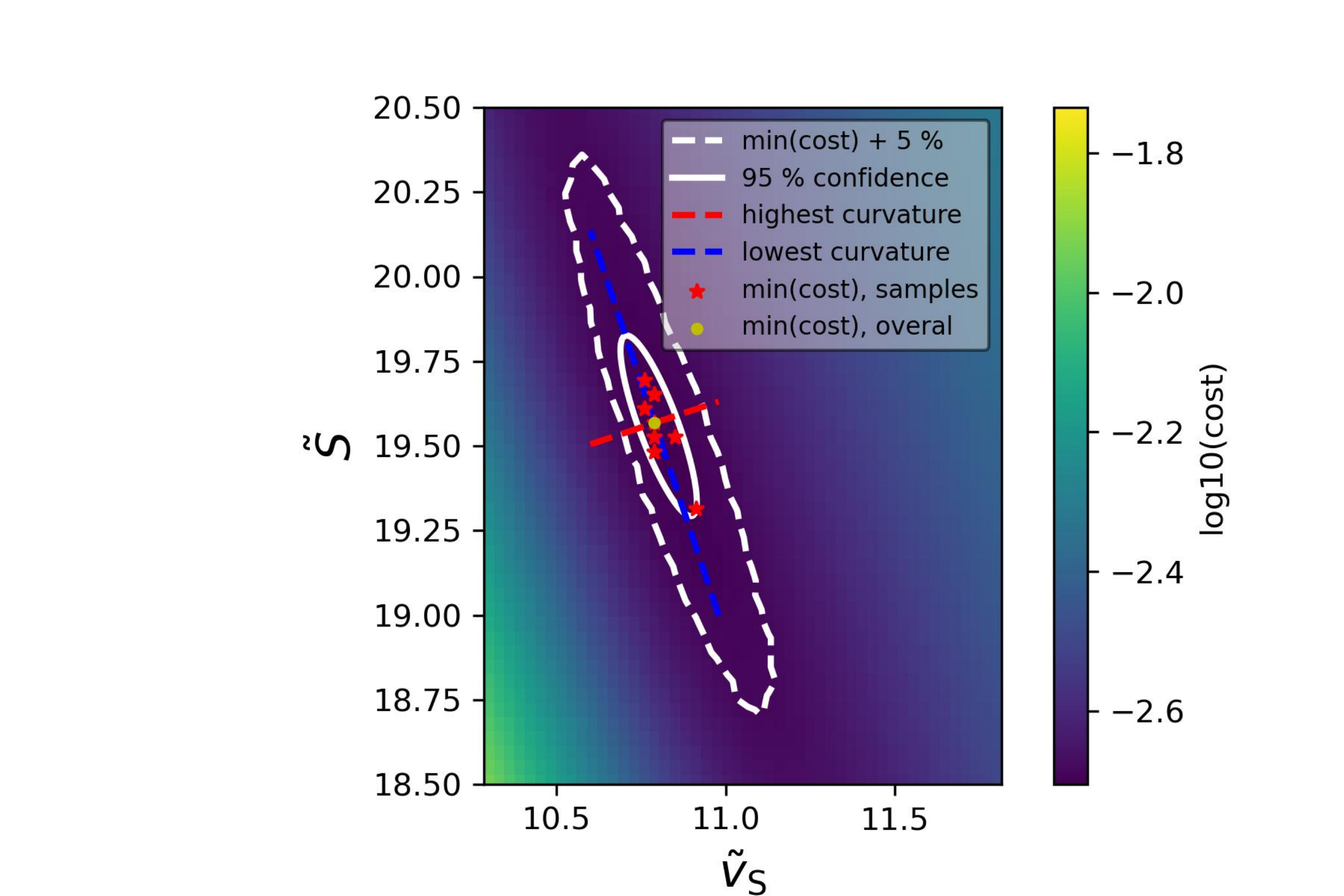}
\caption{
Dependence of the cost function  in Eq.~\ref{eq:cost_function_append} on the independent parameters \(\tilde{v}_\mathrm{S}\) and \(\tilde{S}\). The heatmap shows \(\log_{10}(L(\tilde{S}, \tilde{v}_\mathrm{S}))\) as a function of these parameters, with the color scale indicated by the color bar. The step sizes for \(\tilde{v}_\mathrm{S}\) and \(\tilde{S}\) are 0.0306 and 0.0424, respectively. For each parameter pair, 400 initial cell numbers were independently drawn from the synthetic initial size distributions for mixed organoids, and the average cost was computed. The yellow dot marks the location of the global minimum cost. 
The white dashed contour line indicates the region where the cost increases by 5\% relative to the minimum. Red and blue dashed lines represent the directions of highest and lowest curvature in the cost landscape, respectively. Red stars denote independently estimated minimum-cost locations, obtained from subsets (blocks) of the data. The solid ellipse shows the 95\% confidence region for the estimated position of the global minimum. }
\label{fig:cost_fitting_heatmap}
\end{figure}

In this section, we present an analysis on  sensitivity of the model to the independent parameters. We first verified that we had minimized the cost function~\ref{eq:cost_function_append} by studying its dependence on the parameters $\tilde{v}_S$ and $\tilde{S}$. 
Specifically, we varied \(\tilde{v}_\mathrm{S}\) from 10.3 to 11.8 and \(\tilde{S}\) from 18.5 to 20.5, using step sizes of \(d\tilde{v}_\mathrm{S} = 0.0306\) and \(d\tilde{S} = 0.0424\), respectively. 
These intervals cover at least \(\pm4.6\%\) of \(\tilde{v}_\mathrm{S}\), and  \(\pm4.7\%\) of \(\tilde{S}\) at the point of the minimum cost.
For each parameter pair, we computed the average cost according to Eq.~\ref{eq:cost_function_append} over all the \textit{in-silico} mixed organoids. A total of 700 random initial cell sizes were generated per parameter pair, using the fitted distributions for initial sizes. The average cost was then calculated across the full set of samples, as well as within 14 equal-sized subsets (blocks) of 50 samples each. 
The full set of data points was used to identify the location of the global minimum, while the subsets served to generate independent samples of that location. Figure~\ref{fig:cost_fitting_heatmap} shows the heatmap of the calculated cost function, along with the location of the minimum and its independent samples. To perform a sensitivity analysis of the cost function, we evaluated the Hessian matrix of it by fitting a weighted paraboloid of the form 
\(L\approx a_1 \tilde{v}^2_{\mathrm{S}} 
+ a_2 \tilde{S}^2
+ a_3 \tilde{v}_{\mathrm{S}} \tilde{S}
+ a_4 \tilde{v}_{\mathrm{S}}
+ a_5 \tilde{S}
+ a_6\)
to a \(9 \times 9\) grid of data points centered around the identified global minimum. The second-order partial derivatives obtained from this fit define the entries of the Hessian matrix \(\mathbf{H}_L\), which reads:
\begin{equation}
\mathbf{H}_L =
\begin{bmatrix}
\displaystyle \frac{\partial^2 L}{\partial \tilde{v}_{\mathrm{S}}^2} & \displaystyle \frac{\partial^2 L}{\partial \tilde{v}_{\mathrm{S}} \, \partial \tilde{S}} \\[10pt]
\displaystyle \frac{\partial^2 L}{\partial \tilde{S} \, \partial \tilde{v}_{\mathrm{S}}} & \displaystyle \frac{\partial^2 L}{\partial \tilde{S}^2}
\end{bmatrix}
=
\begin{bmatrix}
\displaystyle 2a_1 & a_3 \\[10pt]
\displaystyle a_3 & 2a_2
\end{bmatrix}
=
\begin{bmatrix}
\displaystyle 0.0104 & 0.0034 \\[10pt]
\displaystyle 0.0034 & 0.0014
\end{bmatrix}.
\label{eq:hessian_mat}
\end{equation}
Since the determinant and the trace of  \(\mathbf{H}_L\) are both positive, the captured critical point is a minimum for the cost function. Moreover, the eigenvectors of the Hessian matrix define the directions for the highest and the lowest increase in the cost function when deviating from the minimum. For the \(\mathbf{H}_L\) above, the eigenvalues and the eigenvectors are 
\(\lambda_1 = 1.15\times10^{-2}\) and \( \lambda_2 = 2.75\times10^{-4}\), with eigenvectors \(\mathbf{v}_1 =[0.95, 0.31]\) and \(\mathbf{v}_2 =[-0.31, 0.95]\), the first and the second components being in the direction of \(\tilde{v}_\mathrm{S}\) and \(\tilde{S}\), respectively. The directions for the steepest and the lowest variations are shown in Fig.~\ref{fig:cost_fitting_heatmap} with red and blue dashed lines.

Moreover, we used two different methods to determine uncertainty in localization of the minimum-cost point. First, we identified the region in which the cost function varies by less than 5\% from its minimum value. This region is shown by the white dashed line in Fig.~\ref{fig:cost_fitting_heatmap}. As evident in the figure, the region forms an elongated ellipse tilted along the direction of lowest curvature of the cost function. This manifests as anisotropy in the sensitivity of the cost function with respect to variations in $\tilde{v}_\mathrm{S}$ and $\tilde{S}$. Within this region, each parameter changes by at most approximately 4\% from its fitted value.
As a second approach, we used independent samples of the minimum-cost location to estimate a 95\% confidence region. Assuming that the variability in the sampled locations arises from a bivariate normal distribution, this confidence region is represented by an ellipse centered at the mean of the samples. Its shape and orientation are determined by the covariance matrix of the samples. The boundary of the ellipse corresponds to points whose squared Mahalanobis distance from the mean equals the 95th percentile of the chi-square distribution with two degrees of freedom, which is $\chi^2_2 \approx 5.99$. This region is shown with a solid white line in Fig.~\ref{fig:cost_fitting_heatmap}.

\end{document}